\documentclass[a4paper,11pt]{article}
\usepackage{graphicx,amssymb,amstext,amsmath}
\usepackage{color}

\newcommand{\bc}{\begin{center}}
\newcommand{\ec}{\end{center}}
\def\ba#1{\begin{array}{#1}\displaystyle}
\newcommand{\ea}{\end{array}}
\newcommand{\z}{\\[2mm] \displaystyle}

\newcommand{\beq}{\begin{equation}}
\newcommand{\eeq}{\end{equation}}
\newcommand{\beqa}{\begin{eqnarray}}
\newcommand{\eeqa}{\end{eqnarray}}
\newcommand{\no}{\nonumber}
\newcommand{\n}{\nonumber\\}
\newcommand{\bi}{\begin{itemize}}
\newcommand{\ei}{\end{itemize}}

\def\lt#1{\left#1}
\def\rt#1{\right#1}
\def\t#1{\tilde{#1}}
\def\h#1{\hat{#1}}
\def\b#1{\bar{#1}}
\def\frc#1#2{\frac{#1}{#2}}

\newcommand{\p}{\partial}

\newcommand{\vac}{{\rm vac}}
\newcommand{\bra}{\langle}
\newcommand{\ket}{\rangle}
\newcommand{\Z}{{\mathbb{Z}}}

\newcommand{\R}{{\mathbb{R}}}
\newcommand{\C}{{\mathbb{C}}}

\newcommand{\Or}{{\cal O}}

\newcommand{\ep}{\epsilon}

\newcommand{\Tr}{{\rm Tr}}

\newcommand{\qi}{{\,\h{\text{\i}}\,}}
\newcommand{\qj}{{\,\h{\text{\j}}\,}}
\newcommand{\qk}{{\,\h{\rm k}\,}}

\date{March 2011}
\title{Correlation functions of twist fields from Ward identities in the massive Dirac theory}
\author{Benjamin Doyon \and James Silk}

\begin{document}

\maketitle

\abstract{We derive non-linear differential equations for correlation functions of $U(1)$ twist fields in the two-dimensional massive Dirac theory. Primary $U(1)$ twist fields correspond to exponential fields in the sine-Gordon model at the free-fermion point, and it is well-known that their vacuum two-point functions are determined by integrable differential equations. We extend part of this result to more general quantum states (pure or mixed) and to certain descendents, showing that some two-point functions are determined by the sinh-Gordon differential equations whenever there is translation and parity invariance, and the density matrix is the exponential of a bilinear expression in fermions. We use methods involving Ward identities associated to the copy-rotation symmetry in a model with two independent, anti-commuting copies. Such methods were used in the context of the thermally perturbed Ising quantum field theory model. We show that they are applicable to the Dirac theory as well, and we suggest that they are likely to have a much wider applicability to free fermion models in general. Finally, we note that our form-factor study of descendents twist fields combined with a CFT analysis provides a new way of evaluating vacuum expectation values of primary $U(1)$ twist fields: by deriving and solving a recursion relation.}

\tableofcontents

\section{Introduction}

The correlation functions of local fields in a quantum field theory (QFT) yield all the physical information contained within a model. In most models these functions are only accessible perturbatively and the expressions obtained only valid for large energies. It is well known, however, that there exists a set of two-dimensional integrable quantum field theories, having free-fermion representations, where non-trivial vacuum correlation functions can be expressed as solutions to integrable differential equations, the first result of this nature being the two-point spin-spin correlation function in the thermally perturbed lattice Ising model \cite{WMTB75}. Since then it has been found that this result, as well as similar results in the QFT model of Dirac fermions, can be retrieved through alternative methods: holonomic quantum fields (Dirac model) \cite{MSJ78}, Fredholm determinants (Ising and Dirac models) \cite{BB92,KBI93,BL97}, determinants of Dirac operators (Dirac model in flat and curved space and with magnetic field) \cite{P90,PBT94,L08}, and doubling of the model (Ising spin chain, and Ising QFT model at zero and non-zero temperature and in curved space) \cite{P80,FZ03,DF04}. The existence of such differential equations is extremely useful, as correlation functions can be evaluated with very high accuracy once the initial conditions have been fixed using conformal perturbation theory and form-factor analysis \cite{D03,DF04,DG08,L08,L09}. Note that, at zero temperature, this provides solutions to non-trivial connection problems in the theory of Painlev\'e equations.

Thanks in great part to the holonomic quantum field description, it is understood that such differential equations should exist in general for correlation functions of twist fields in free fermion models. Twist fields are local fields which have, with respect to the fermion fields, non-trivial monodromy associated to a symmetry transformation. The concept of twist fields was first introduced in \cite{ST78}, as the $\Z_2$ monodromy field of the Majorana fermion corresponding to the spin field of the Ising model. Similarly, it is known that primary $U(1)$ twist fields in the Dirac model reproduce correlation functions of exponential fields in the sine-Gordon model at a particular value of the coupling (the free-fermion point). It is correlation functions of such $\Z_2$ and $U(1)$ primary twist fields that have been studied and that are known to lead to differential equations in QFT models.

In this paper, we examine the model of free massive Dirac fermions and study two-point correlation functions of $U(1)$ twist fields. The method used to study the correlation functions is similar to that of \cite{FZ03} where differential equations for Ising spin-spin correlation functions are found by looking at conserved charges in a model containing two non-interacting copies of the Ising field theory. The associated Ward identities naturally give a bilinear form of the differential equations. However, in \cite{FZ03}, many particularities of the Ising model were used to obtain such results. The aim of this paper is four-fold: to clarify the method of \cite{FZ03} and show how it works beyond the Ising model and beyond the finite-temperature density matrix (suggesting that it should work more generally for any free-fermion model and quite general density matrices expressed as exponentials of bilinears); to establish more precisely in the Dirac theory differential equations for twist-field correlation functions in general quantum states, pure or mixed (under the condition that the density matrix is the exponential of a bilinear expression in the fermions); to establish equations determining two-point functions involving certain descendent fields; and to derive a recursion relation for vacuum expectation values (VEVs) of primary twist fields which can be used to evaluate them.

The method of \cite{FZ03} was widely thought to be only applicable to the Ising model because of the particular form of the OPEs: only the primary bosonic twist field, its unique fermionic descendent, and their derivatives are involved. In the Dirac theory, there are many more fermionic descendents, and there are bosonic descendents that are not derivatives of primaries. Yet, the method still works. We believe that the method should work quite generally because it is intimately related to standard techniques used in the context of classical integrable models, where bilinear differential equations are naturally obtained. The method of \cite{FZ03} was also slightly restricted in other ways: it only dealt with the vacuum or the finite-temperature states, and it used explicitly translation invariance in order to construct particular conserved charges and derive differential equations from the associated Ward identities. We show, in the context of the Dirac theory, that much more general quantum states can be used, and we show how to obtain equations even without translation invariance, only using a single conserved charge (something which was initially noticed, in the Ising model, in the thesis of one of the authors \cite{Dth}). We note, however, that only when the quantum state is translation and parity invariant, these equations simplify to something that is tractable. Interestingly, this new way clearly separates the origin of the mass dependence on the Ward identities from pure CFT calculations.

In order to give a full understanding of the way the equations are derived, we develop the theory of Dirac $U(1)$ twist fields to some extent: we construct spinless, chargeless twist fields $\sigma_\alpha$ for all non-integer $\alpha\in \R$, characterised by their $U(1)$ monodromy $e^{2\pi i \alpha}$ and dimension $\alpha^2$, as well as their fermionic descendents defined via the operator product expansion (OPE), $\sigma_{\alpha+1,\alpha}\sim \Psi_R^\dag\sigma_\alpha$ and $\sigma_{\alpha-1,\alpha}\sim\Psi_R\sigma_\alpha$ (where $\Psi_R$ is the right-moving component of the Dirac spinor). The construction is done both using explicit form factors, and using a CFT description that takes into account the constraints provided by the massive perturbation. We evaluate the action of the double-model conserved charge using these two pictures.

After calculating the form factors and examining the OPEs, a recursion relation for the VEVs $\bra\vac|\sigma_\alpha|\vac\ket = m^{\alpha^2} c_{\alpha}$ is found:
\begin{equation}
 c_{\alpha}=\frac{\Gamma(\alpha+1)}{\Gamma(-\alpha)}c_{\alpha+1}.
\end{equation}
This relation is consistent with known results \cite{LZ96} but can also be viewed as a method for evaluating $c_{\alpha}$, given the normalisation $c_{0}=1$ and the symmetry $c_{-\alpha} = c_\alpha$

The main results are as follows. Consider correlation functions in a quantum state whose density matrix is an exponential of a bilinear form in fermions, and which is translation and parity invariant. Writing the following correlation function as
\begin{equation}
 \langle\sigma_{\alpha}(x,y)\sigma_{\alpha}(0,0)\rangle= c_\alpha^2 m^{2\alpha^2}e^{\Sigma_{\alpha}(x,y)}
\end{equation}
we show that it can be parametrised by a single function $\psi$ where
\begin{align}
 \partial\bar{\partial}\psi & =\frac{m^{2}}{2}\sinh(2\psi) \label{eq:result1} \\
 \partial\bar{\partial}\Sigma_{\alpha} & =\frac{m^{2}}{2}(1-\cosh(2\psi)). \label{eq:result2}
\end{align}
This result was first found implicitly in \cite{MSJ78} and then explicitly, using the method of Fredholm determinants, in \cite{BL97}. We further find that the correlation function involving the descendent field $\sigma_{\alpha+1}$ along with $\sigma_\alpha$ is obtained from
\begin{equation}
 \langle\vac|\sigma_{\alpha}(x,y)\sigma_{\alpha+1}(0,0)|\vac\rangle=
 c_\alpha c_{\alpha+1} m^{2\alpha^2+2\alpha+1}e^{\Sigma_{\alpha}'(x,y)}
\end{equation}
via
\beq\label{eq:result3}
	 \partial\bar{\partial}\Sigma'_{\alpha}=
	 \frac{2\tanh^{2}(\psi)}{\cosh(2\psi)-1}\partial\psi\bar{\partial}\psi.
\eeq
Note that this is different from the equation found for the correlation functions \\ $\bra\vac|\sigma_\alpha(x,y)\sigma_\beta(0,0)|\vac\ket$ in \cite{BL97}, but there is no contradiction, as there there was the restriction $0<\alpha,\beta<1$. Finally, we find that the correlation function of fermionic descendents,
\beq
	\bra\sigma_{\alpha+1,\alpha}(x,y)\sigma_{\alpha,\alpha+1}(0,0)\ket =
	e^{-i\pi \alpha} c_\alpha c_{\alpha+1} m^{2\alpha^2+2\alpha+1} \Omega(x,y),
\eeq
is algebraically expressed in terms of the previous correlation functions:
\beq\label{eq:result4}
	\Omega = \sqrt{e^{2\Sigma_\alpha} - e^{2\Sigma_\alpha'}}.
\eeq

The paper is set out as follows: in Section \ref{sec:model} notation and conventions are established and the definitions of the primary twist fields are made explicit. In Section \ref{sec:desfields} the descendent twist fields are introduced through the OPEs of fermion fields and primary twist fields. The one- and then multi-particle form factors are calculated and their consistency is checked. A recursion relation for the constants $c_{\alpha}$ is found and discussed. In Section \ref{sec:DoubleMod} the double Dirac model is introduced. Conserved charges are discussed and the method by which the action of these charges is verified is described. In Section \ref{sec:corfun} the appropriate Ward identities are written down and simplified to give the main result. Section \ref{sec:sum} gives our conclusions and possible areas for future work. Finally the appendices contain a discussion of the CFT considerations, and of the Ward identities of Section \ref{sec:corfun} in a model without translation, rotation and parity symmetry.


\section{The model and its primary fields} \label{sec:model}

Throughout this paper, $x\in\R$ is a space coordinate and $y\in\R$ is a Euclidean time coordinate ($y=it$, where $t$ is real time). We will put space and Euclidean time into complex coordinates $z:=-\frac{i}{2}(x+iy)$, $\bar{z}:=\frac{i}{2}(x-iy)$, with derivatives $\partial:=\partial_{z} =i\partial_{x}+\partial_{y}$ and $\bar{\partial}:=\partial_{\bar{z}}=-i\partial_{x}+\partial_{y}$. Also, when only one coordinate is specified for a quantum field, it is understood as the space coordinate, the Euclidean time coordinate being set to 0.


\subsection{Dirac fermions}
In two dimensions and in the quantisation on the line, the free Dirac Fermi field of mass $m$ is an operator solution $\Psi_R(x,y),\;\Psi_L(x,y)$ to the equations of motion
\beq\ba{ll}
 \bar{\partial}\Psi_{R}=-im\Psi_{L},& \bar{\partial}\Psi_{R}^{\dagger}=-im\Psi_{L}^{\dagger} \z
 \partial\Psi_{L}=im\Psi_{R}, & \partial\Psi_{L}^{\dagger}=im\Psi_{R}^{\dagger} \ea \label{eom}
\eeq
and to the equal-time anti-commutation relations
\begin{equation}
	\{ \Psi_{R}(x_{1}),\Psi_{R}^{\dagger}(x_{2})\}=4\pi\delta(x_{1}-x_{2}) ,\quad \{\Psi_{L}(x_{1}),\Psi_{L}^{\dagger}(x_{2})\}=4\pi\delta(x_{1}-x_{2}) \label{eq:diracanticom}
\end{equation}
(other combinations of fields anti-commutating), subject to the condition that there exists a state $|\vac\ket$, the vacuum state, such that
\beq\label{defvac}
	\lim_{y\to-\infty} \Psi_{R,L}(x,y)|\vac\ket =
	\lim_{y\to-\infty} \Psi_{R,L}^\dag(x,y)|\vac\ket = 0.
\eeq
The normalisation in (\ref{eq:diracanticom}) is chosen in such a way that the fields have the so-called CFT normalisation in terms of the $z$ variables: as $|z_1-z_2|\to 0$,
\beqa
	\bra\vac| {\cal T}\lt[\Psi_R^\dag(x_1,y_1) \Psi_R(x_2,y_2)\rt]|\vac\ket &\sim&
		\frc1{z_1-z_2},\n
	\bra\vac| {\cal T}\lt[\Psi_L^\dag(x_1,y_1) \Psi_L(x_2,y_2)\rt]|\vac\ket &\sim& \frc1{\b{z}_1-\b{z}_2}
		.\no
\eeqa
Here, the symbol ${\cal T}$ is the time-ordering symbol: in its bracket, operators at later Euclidean times are placed to the left of operators at earlier Euclidean times (getting a sign, in general, if the operators exchanged have fermionic statistics).

As usual, the solution can be given in terms of creation and annihilation operators (Fourier modes of the fields), whose time evolution is simple:
\beqa
 \Psi_{R}(x,y)&=&\sqrt{m}\int{\rm d}\theta\, e^{\theta/2}\lt(D_{+}^{\dagger}(\theta)e^{y E_\theta-ix p_\theta}-i D_{-}(\theta)e^{-y E_\theta+ix p_\theta}\rt) \nonumber \\
 \Psi_{L}(x,y)&=&\sqrt{m}\int{\rm d}\theta \,e^{-\theta/2}\lt(iD_{+}^{\dagger}(\theta)e^{y E_\theta-ix p_\theta}-D_{-}(\theta)e^{-y E_\theta+ix p_\theta}\rt)\label{psi}
\eeqa
where
\beq
	E_\theta = m\cosh\theta,\quad p_\theta = m\sinh\theta.
\eeq
It is worth noting that since $y$ is a Euclidean time, the Hermitian conjugate operators are obtained from $\Psi_{R,L}$ by direct Hermitian conjugation of the expressions above, followed by a change of sign of $y$:
\beqa
 \Psi_{R}^\dag(x,y)&=&\sqrt{m}\int{\rm d}\theta\, e^{\theta/2}\lt(iD_{-}^{\dagger}(\theta)e^{y E_\theta-ix p_\theta}+D_{+}(\theta)e^{-y E_\theta+ix p_\theta}\rt) \nonumber \\
 \Psi_{L}^\dag(x,y)&=&\sqrt{m}\int{\rm d}\theta \,e^{-\theta/2}\lt(-D_{-}^{\dagger}(\theta)e^{y E_\theta-ix p_\theta}-iD_{+}(\theta)e^{-y E_\theta+ix p_\theta}\rt)\label{psidag}
\eeqa
From (\ref{eq:diracanticom}), the creation and annihilation operators satisfy
\beq
 \{D_{+}^{\dagger}(\theta_1),D_{+}(\theta_2)\}=\delta(\theta_1-\theta_2),\quad \{D_{-}^{\dagger}(\theta_1),D_{-}(\theta_2)\}=\delta(\theta_1-\theta_2) \label{eq:diraccandaanticom}
\eeq
(with all other combinations anti-commuting), and from (\ref{defvac}), the vacuum condition
\[
	D_\pm(\theta)|\vac\ket = 0.
\]
The Hilbert space is simply the Fock space over (\ref{eq:diraccandaanticom}), with a basis of multi-particle states denoted by
\begin{equation}
 |\theta_{1}\cdots\theta_{n}\rangle_{\epsilon_{1}\cdots\epsilon_{n}}:=D_{\epsilon_{1}}^{\dagger}(\theta_{1})\cdots D_{\epsilon_{n}}^{\dagger}(\theta_{n})|\vac\rangle \quad \mbox{for}\quad \theta_1>\cdots>\theta_n,\ \epsilon_{i}\in\{+,-\}. \label{basis}
\end{equation}
These states have total energies $\sum_j E_{\theta_j}$ and total momenta $\sum_j p_{\theta_j}$. The dual vectors will be denoted by ${}^{\epsilon_{1}\cdots\epsilon_{n}}\bra\theta_{1}\cdots\theta_{n}|:= |\theta_{1}\cdots\theta_{n}\rangle_{\epsilon_{1}\cdots\epsilon_{n}}^\dag$, and clearly we have
\begin{equation}
	{}^{\epsilon_{1}'\cdots\epsilon_{n}'}\bra\theta_{1}'\cdots\theta_{n}'|
	\theta_{1}\cdots\theta_{n}\rangle_{\epsilon_{1}\cdots\epsilon_{n}}
	=\prod_{j=1}^n \delta^{\epsilon_j'}_{\epsilon_j}\delta(\theta_j-\theta_j^{\prime}).
\end{equation}
Writing as in (\ref{basis}) vectors with different orderings of the creation operators, the identity operator can be decomposed into
\begin{equation}
 1=\displaystyle\sum_{N=0}^{\infty}\frac{1}{N!}\displaystyle\sum_{\epsilon_{1}\cdots\epsilon_{N}}\int_{-\infty}^{\infty}d\theta_{1}\cdots\int_{-\infty}^{\infty}d\theta_{N}\,|\theta_{1}\cdots\theta_{N}\rangle_{\epsilon_{1}\cdots\epsilon_{N}}^{\phantom{\epsilon_{1}\cdots\epsilon_{N}}\epsilon_{N}\cdots\epsilon_{1}}\langle\theta_{N}\cdots\theta_{1}|. \label{eq:resofid}
\end{equation}

The solution (\ref{psi}), (\ref{psidag}), defining the mode operators from the Fermion operators, is chosen in such a way that it satisfies the equations of motion, and that the resulting mode operators obey the canonical anti-commutation relations (\ref{eq:diraccandaanticom}). However, the choice of such mode operators is not unique: we may re-define the mode operators via $D_\pm(\theta)\mapsto u_\pm D_\pm(\theta)$ for some pure phases $u_\pm$, $|u_\pm|=1$, and still have a solution to the equations of motion and canonical anti-commutation relations. This amounts to a phase-redefinition of the states in the Hilbert space. The mode operators are partially fixed by requiring further the crossing symmetry condition
\beq\label{crosssym}
	\bra\vac|\Psi_{R,L}(0)|\theta+i\pi\ket_\ep = {}^{-\ep}\bra\theta|\Psi_{R,L}(0)|\vac\ket.
\eeq
This condition leaves one phase ambiguity: there is still an invariance under the change $D_\pm(\theta)\mapsto u^{\pm1} D_\pm(\theta)$ for $|u|=1$. This phase ambiguity will not play any role in the following, hence we fix it arbitrarily by choosing the solution (\ref{psi}), (\ref{psidag}).


\subsection{Primary twist fields}

The Dirac theory enjoys a $U(1)$ internal symmetry: $\Psi_{R,L}\mapsto e^{2\pi i\alpha} \Psi_{R,L}$, $\alpha\in[0,1)$. The associated primary (with respect to the Fermion algebra) twist fields, $\sigma_{\alpha}(x,y)$, are local, spin-less, $U(1)$ neutral quantum fields with bosonic statistics, and are known to have scaling dimensions $\alpha^{2}$ \cite{MSJ78}. Their twist property, characterised by a locality index $e^{2\pi i\alpha}$, can be expressed via a monodromy property of correlation functions of time-ordered operators,
\[
	C(z) := \bra\vac|{\cal T}\lt[\cdots \Psi_{R,L}(x,y)\sigma_{\alpha}(0)\cdots\rt] |\vac\ket.
\]
Seeing $C(z)$ as a function of the complex variable $z$ smoothly continued through the $y=0$ line, its continuation along a loop $\gamma$ surrounding once counter-clockwise the origin is given by
\begin{equation}
 C(e^{2\pi i}z) =e^{-2\pi i\alpha} C(z).
\end{equation}
This is true for any loop $\gamma$ that can be contracted to the origin without intersecting the position of other twist fields possibly present in the correlation function. A similar relation holds for the Hermitian conjugates $\Psi_{R,L}^\dag$, but with the factor $e^{2\pi i\alpha}$ on the right-hand side, instead of $e^{-2\pi i\alpha}$. This twist property can also be expressed operatorially via equal-time exchange relations:
\begin{subequations} \label{eq:equaltimebraiding}
\beqa
 \Psi_{R,L}(x)\sigma_{\alpha}(0)&=&\left\{ \begin{array}{ll}
                                         \sigma_{\alpha}(0)\Psi_{R,L}(x) & (x<0)  \z
					 e^{2\pi i\alpha}\sigma_{\alpha}(0)\Psi_{R,L}(x) & (x>0)
                                        \end{array} \right.  \\
 \Psi_{R,L}^{\dagger}(x)\sigma_{\alpha}(0)&=&\left\{ \begin{array}{ll}
                                         \sigma_{\alpha}(0)\Psi_{R,L}^{\dagger}(x) & (x<0)  \z
					 e^{-2\pi i\alpha}\sigma_{\alpha}(0)\Psi_{R,L}^{\dagger}(x) & (x>0).
                                        \end{array} \right.
\eeqa
\end{subequations}
These naturally lead us to define primary twist fields with negative index via Hermitian conjugates:
\beq\label{herm}
	\sigma_{-\alpha} := \sigma_\alpha^\dag,\quad \alpha\in[0,1).
\eeq

These exchange relations along with the property of being primary immediately gives the form factors of the $U(1)$ twist fields \cite{K78,MSS81,EL93}. As the fields are $U(1)$ neutral they only have non-vanishing form factors with $U(1)$ neutral states, so for $-1<\alpha<1$,
\begin{eqnarray} \label{eq:Formfactorsaa}
 \langle \vac|\sigma_{\alpha}(0)|\theta_{1}\theta_{2}\cdots\theta_{2n}\rangle_{+\cdots+-\cdots-}  = & c_{\alpha}m^{\alpha^{2}}(-1)^{n(n-1)/2}\left(\frac{\sin(\pi\alpha)}{\pi i}\right)^{n}\left(\prod^{n}_{i=1}(u_{i})^{\frac{1}{2}+\alpha}(u_{i+n})^{\frac{1}{2}-\alpha}\right) \nonumber \\
  & \times\ \frac{\prod_{i<j\leq n}(u_{i}-u_{j})\prod_{n+1\leq i<j}(u_{i}-u_{j})}{\prod_{r=1}^{n}\prod_{s=n+1}^{2n}(u_{r}+u_{s})}
 	 \label{eq:sigmaaadef}
\end{eqnarray}
where there are $n$ $-$'s and $n$ $+$'s in the state, and $u_{i}:=\exp(\theta_{i})$. In particular, the two-particle form factor is
\beq\label{2pff}
	\bra\vac|\sigma_\alpha(0)|\theta_1\theta_2\ket_{+-} = c_\alpha m^{\alpha^2}
	\frc{\sin(\pi\alpha)}{2\pi i} \frc{e^{\alpha(\theta_1-\theta_2)}}{\cosh\frc{\theta_1-\theta_2}2}.
\eeq
Note that the form factors (\ref{eq:Formfactorsaa}) may differ from those written in other publications in the choice of the sign of $\alpha$ and in the choice of the sign of the two-particle form factor (the latter amounting to an overall sign $(-1)^n$). These choices relate to our choice of exchange relations (\ref{eq:equaltimebraiding}) (defining the operator $\sigma_\alpha$), and to our choice of crossing symmetry relation (\ref{crosssym}) (defining the asymptotic states). Calculations below, using the two-particle form factor, will show that the two-particle form factor (\ref{2pff}) leads to a behaviour of $\Psi_R^\dag(x)\sigma_\alpha(0)$ and $\Psi_R(x)\sigma_\alpha(0)$ as $x\to0$ exhibiting a branch point in accordance to the exchange relations (\ref{eq:equaltimebraiding}). Further calculations will show that the disconnected, delta-function term in a crossing symmetry relation similar to (\ref{crosssym}) is correct, this being rue only with the sign chosen for the two-particle form factor.

The constant $c_\alpha$ is chosen, up to a phase, in order to guarantee the CFT normalisation of the fields:
\beq\label{normsigm}
	\bra\vac|\sigma_\alpha(x_1,y_1)\sigma_{-\alpha}(x_2,y_2)|\vac\ket
		\sim \frc1{|z_1-z_2|^{2\alpha^2}}\quad\mbox{as}\quad |z_1-z_2|\to0.
\eeq
The phase of $c_\alpha$ is fixed by requiring that it be real and positive. The evaluation of the constant $c_\alpha$ was first explained in \cite{LZ96} (this constant was first obtained in an unpublished note by Al.B. Zamolodchikov). This constant, or rather its generalisation to the Dirac Fermi field on the Poinar\'e disk, was identified with Barnes' $G$-function in \cite{D03}:
\beq\label{calpha}
	c_\alpha = \frc1{G(1-\alpha)G(1+\alpha)}.
\eeq
Recall that the main property of Barnes' $G$-function is $G(1+z)=\Gamma(z)G(z)$, with the normalisation $G(1) = 1$. In the next section we provide an alternative way of evaluating $c_\alpha$.

Other matrix elements can be obtained by analytic continuation in the rapidities, as is standard in the context of 1+1-dimensional QFT \cite{S92}. Note that the form factors (\ref{eq:sigmaaadef}) and the Hermiticity relation (\ref{herm}) are in agreement with the analytic-continuation formula
\beq
	\langle \vac|\sigma_{\alpha}(0)|(\theta_{1}+i\pi)\cdots(\theta_{2n}+i\pi)\rangle_{+\cdots+-\cdots-}
	={}^{+\cdots+-\cdots-}\bra \theta_{2n} \cdots \theta_1|\sigma_{\alpha}(0)|\vac\ket,
\eeq
where the analytic continuation is simultaneous on all rapidities.


\section{Descendent twist fields and their form factors} \label{sec:desfields}

We are interested in deriving differential equations for two-point correlation functions of the primary twist fields $\sigma_\alpha$. However, in the derivation, the intermediate steps naturally involve other twist fields; descendents with respect to the Fermion algebra. Here we provide unambiguous definitions for these fields and evaluate their form factors.

\subsection{Basic definitions}

First, notice that (\ref{eq:sigmaaadef}) is an analytic function of $\alpha$ on $\C\setminus \Z^*$, for fixed rapidities; the points $\Z^*:= \Z\setminus \{0\}$ correspond in general to poles. Hence, it is natural to define the fields $\sigma_\alpha$ for all $\alpha\in\R\setminus \Z^*$ by formula (\ref{eq:sigmaaadef}) (and by analytic continuation in the rapidities for other matrix elements). Then, the two-point function of such twist fields is, for any fixed distance between the fields, an analytic function of $\alpha$ on a neighbourhood of $\R\setminus \Z^*$. In fact, we expect that the coefficients of operator product expansions with the stress-energy tensor are ``uniformly'' analytic in $\alpha$ at all distances. This means that the new field defined by analytic continuation has dimension $\alpha^2$; we also expect the correct CFT normalisation (\ref{normsigm}).

Other descendent twist fields, of Fermionic statistics (in particular, whose form factors with even number of particles are zero), can be obtained as regularised limits where the position of a Dirac field approaches that of a twist field $\sigma_\alpha$; equivalently as coefficients in the operator product expansions (OPEs). For $\alpha$ in particular subsets of $\R\setminus \Z^*$, we define two one-parameter families of Fermionic twist fields via the coefficients occurring in the leading short-distance asymptotics associated to the Dirac fields $\Psi_{R}$ and $\Psi_{R}^\dag$:
\begin{align}
 {\cal T}[\Psi_{R}^{\dagger}(x,y)\sigma_{\alpha}(0)]&\sim(-iz)^{\alpha}\sigma_{\alpha+1,\alpha}(0) && \text{for}~ \alpha<\frc12 \n
 {\cal T}[\Psi_{R}(x,y)\sigma_{\alpha}(0)]&\sim(-iz)^{-\alpha}\sigma_{\alpha-1,\alpha}(0) && \text{for}~ \alpha>-\frc12. \label{ftw}
\end{align}
As above, ${\cal T}$ is the time-ordering operation. The power functions on the right-hand sides are on their principal branches; note that there is agreement between the phases occurring from exchange relations (\ref{eq:equaltimebraiding}) and the phase differences occurring through the cuts of the principal branches. The requirement of phase agreement is not quite enough to fully determine the particular power functions occurring: these are consequences of CFT considerations (see appendix \ref{sec:cft}), or of form-factor calculations (performed below). From (\ref{ftw}), we see that the new fields $\sigma_{\alpha\pm 1,\alpha}$ have dimensions $\alpha^2\pm\alpha+1/2$, spins $\pm \alpha+\frc12$ and charges $\mp 1$. They also satisfy the hermiticity relation
\beq\label{hermdesc}
	\sigma_{\alpha\pm1,\alpha}^\dag = \sigma_{-\alpha\mp1,-\alpha}.
\eeq
The range of $\alpha$ shown is a consequence of the equations of motion, as we explain shortly. We define fields for all $\alpha\in\R\setminus\Z^*$ again by analytic continuation beyond the range of $\alpha$ shown.

CFT considerations (appendix \ref{sec:cft}) also suggest that we may define the same Fermionic twist field families using $\Psi_L$ and $\Psi_L^\dag$:
\begin{align}
 {\cal T}\lt[\Psi_{L}^{\dagger}(x,y)\sigma_{\alpha}(0)\rt]&\sim-(i\bar{z})^{-\alpha}\sigma_{\alpha,\alpha-1}(0) &&
 	\text{for}~ \alpha>-\frc12 \n
 {\cal T}\lt[\Psi_{L}(x,y)\sigma_{\alpha}(0)\rt]&\sim-(i\bar{z})^{\alpha}\sigma_{\alpha,\alpha+1}(0) &&
 	\text{for}~ \alpha<\frc12. \label{consftw}
\end{align}
Together with the equations of motion (\ref{eom}), this suggests the form of the leading OPE for all $\alpha\in\R\setminus \Z^*$:
\beqa
 {\cal T}\lt[\Psi_{R}^{\dagger}(x,y)\sigma_{\alpha}(0)\rt]&\sim&(-iz)^{\alpha}\sigma_{\alpha+1,\alpha}(0) 
 	+ \frc{m}{1-\alpha} (i\b{z})^{1-\alpha}\sigma_{\alpha,\alpha-1}(0)\n
 {\cal T}\lt[\Psi_{R}(x,y)\sigma_{\alpha}(0)\rt]&\sim&(-iz)^{-\alpha}\sigma_{\alpha-1,\alpha}(0)
 	+ \frc{m}{1+\alpha} (i\b{z})^{1+\alpha}\sigma_{\alpha,\alpha+1}(0) \label{ftwfull}
\eeqa
and
\beqa
 {\cal T}\lt[\Psi_{L}^{\dagger}(x,y)\sigma_{\alpha}(0)\rt]&\sim&-(i\bar{z})^{-\alpha}\sigma_{\alpha,\alpha-1}(0)
 	- \frc{m}{1+\alpha} (-iz)^{1+\alpha}\sigma_{\alpha+1,\alpha}(0)  \n
 {\cal T}\lt[\Psi_{L}(x,y)\sigma_{\alpha}(0)\rt]&\sim&-(i\bar{z})^{\alpha}\sigma_{\alpha,\alpha+1}(0)
 	- \frc{m}{1-\alpha} (-iz)^{1-\alpha}\sigma_{\alpha-1,\alpha}(0), \label{consftwfull}
\eeqa
where the meaning of $\sim$ is that for any value of $\alpha$, the term on the right-hand side that is dominant gives the leading OPE. Note that according to these equations, the leading term of the OPE occurring in the CFT is modified in certain ranges of $\alpha$ by lower-dimensional ``ghost fields'', like $m\sigma_{\alpha,\alpha-1}$, which make no contribution in the massless limit. Note also that sub-leading terms in these OPE's will have poles at other integer values of $\alpha$.

Moreover, further applications of Dirac fields may lead back to the original family $\sigma_\alpha$ of twist fields. For instance, we have, for $\alpha<0$,
\beq\label{ts}
	{\cal T}\lt[\Psi_R^\dag(x,y)\sigma_{\alpha,\alpha+1}(0)\rt] \sim -i(-iz)^\alpha \sigma_{\alpha+1}(0).
\eeq
This relation is an easy consequence of CFT considerations (see appendix \ref{sec:cft}). Note in particular that the double application of Dirac fields preserved the CFT normalisation, because the Dirac fields themselves are CFT normalised. 

The next two subsections are devoted to computing the form factors of the Fermionic twist fields defined in (\ref{ftw}), and to showing the correctness of (\ref{ftwfull}) and (\ref{consftwfull}). In combination with these form factors, the consistency of (\ref{ftw}) and (\ref{consftw}) leads to the non-trivial recursion relation (\ref{eq:rationofc}) for the normalisation constant $c_\alpha$. Equivalently, relation (\ref{ts}) can be used (since (\ref{ts}) is a simpler consequence of CFT, this is a slightly more direct way of obtaining the recursion relation (\ref{eq:rationofc})). We will show that it is indeed satisfied by (\ref{calpha}). Interestingly, such a recursion relation can be seen as a novel way of evaluating this constant.


\subsection{One particle form factors} \label{sec:1pff}
In this subsection we give details of the computation of $\langle\vac|\sigma_{\alpha+1,\alpha}(0)|\theta\rangle_{+}$. The form factor is evaluated by expanding  $\langle\vac|\Psi_{R}^{\dagger}(x)\sigma_{\alpha}(0)|\theta\rangle_{+}$ and finding the term proportional to $z^{\alpha}$, using (\ref{ftw}). We provide the result for the form factor $\bra\vac|\sigma_{\alpha-1,\alpha}(0)|\theta\ket_-$, which was obtained by expanding $\bra\vac|\Psi_R(x)\sigma_\alpha(0)|\theta\ket_-$ similarly.

In order to obtain the leading short-distance behaviour of
\begin{equation}
\langle\vac|\Psi_{R}^{\dagger}(x,y)\sigma_{\alpha}(0)|\theta\rangle_{+}
\end{equation}
we evaluate it by inserting (\ref{eq:resofid}) between the two fields. We consider $x<0,\;y=0$ for simplicity, as this is sufficient to extract the one-particle form factors. The full $z$ dependence of the leading behaviour can be restored using rotation (or relativistic boost) covariance. There is only one non-zero form factor of $\Psi_{R}^{\dagger}$ given by
\beq
 \langle\vac|\Psi_{R}^{\dagger}(x)|\theta\rangle_{+}=  \sqrt{m}e^{\theta/2}e^{ixp_{\theta}} \label{eq:1pff}
\eeq
This means that there is only one term involving $\sigma_{\alpha}$ to be evaluated, which can be accomplished using the form factor rules of \cite{S92} and the two-particle form factor (\ref{2pff}):
\begin{align}
 ^{+}\langle\phi|\sigma_{\alpha}(0)|\theta\rangle_{+}&= \langle\vac|\sigma_{\alpha}(0)|\theta,\phi+i\pi-i0^+\rangle_{+-}+\bra\vac|\sigma_\alpha(0)|\vac\ket\delta(\theta-\phi) \nonumber \\
 &= c_{\alpha}m^{\alpha^{2}}\frac{\sin (\pi\alpha)}{2\pi}e^{-i\pi\alpha} \frac{e^{\alpha(\theta-\phi)}}{\sinh \lt(\frac{\theta-\phi+i0^+}{2}\rt)}+c_{\alpha}m^{\alpha^2}\delta(\theta-\phi). \label{eq:rotatedpart}
\end{align}
The $+i0^+$ prescription is to be understood in terms of distributions: it is a prescription as to how an integral should avoid the pole of the function. Putting these together we see that
\beqa
 \langle\vac|\Psi_{R}^{\dagger}(x)\sigma_{\alpha}(0)|\theta\rangle_{+}&=& c_{\alpha}m^{\alpha^{2}+1/2}\frac{\sin(\pi\alpha)}{2\pi}e^{-i\pi\alpha}\int{\rm d}\phi\,e^{\phi/2}\frac{e^{\alpha(\theta-\phi)}}{\sinh\lt(\frac{\theta-\phi+i0^+}{2}\rt)}e^{ixp_{\phi}} \n && +\; c_{\alpha}m^{\alpha^2+1/2} e^{\theta/2} e^{ixp_\theta}.
\eeqa
The integral is conditionally convergent. In order to make it a convergent integral, we make a change of contour, shifting $\phi\mapsto\phi-i\pi/2$. This shift of contour does not give any pole contribution thanks to the $i0^{+}$ prescription, and the result is indeed a convergent integral for $x<0$:
\beqa
 \langle\vac|\Psi_{R}^{\dagger}(x,0)\sigma_{\alpha}(0)|\theta\rangle_{+} =&  c_{\alpha}m^{\alpha^{2}+1/2}\frac{\sin(\pi\alpha)}{2\pi}e^{-i\pi\alpha/2}\int{\rm d}\phi\,e^{-i\pi/4}e^{\phi/2}\frac{e^{\alpha(\theta-\phi)}}{\sinh(\frac{\theta-\phi}{2}+\frac{i\pi}{4})}e^{xE_{\phi}} \n & +\; c_{\alpha}m^{\alpha^2+1/2} e^{\theta/2} e^{ixp_\theta} \n
  = & c_{\alpha}m^{\alpha^{2}+1/2}\frac{\sin(\pi\alpha)}{\pi}e^{-i\pi\alpha/2}\int{\rm d}\phi\, \frac{e^{\alpha\theta-(\alpha-1/2)\phi+xE_{\phi}}} {i\,e^{\frac{\theta-\phi}{2}}-e^{\frac{\phi-\theta}{2}}}
\n & +\; c_{\alpha}m^{\alpha^2+1/2} e^{\theta/2} e^{ixp_\theta}.\label{integral}
\eeqa

In general, the leading behaviour at small $x$ will be obtained from the large-$\phi$ behaviour of the integrand. Let us analyse what happen at $x=0$ in different regions of $\alpha$. In the regions $\alpha<0$ and $\alpha>1$, the resulting integral is divergent. In these cases, the leading small-$x$ behaviour is growing, and can be obtained from the leading large-$|\phi|$ behaviour of the integrand only. On the other hand, in the region $0<\alpha<1$, the integral converges, and the resulting constant is cancelled by the term arising from the delta function (see Subsection \ref{ssrec}). Note that this cancellation requires in particular the correct sign of the two-particle form factor (\ref{2pff}). Hence, in this case, the leading small-$x$ behaviour is decaying, and can be extracted again from the leading large-$|\phi|$ behaviour by first rendering the integral divergent by taking its first $x$-derivative. As explained before, we expect the field $\sigma_{\alpha+1,\alpha}$ to occur in the leading OPE for $\alpha<1/2$, and to have an appropriate ghost fields in the range $\alpha>1/2$, as per (\ref{ftwfull}). We will see below that the one-particle form factors agree with this. In order to evaluate the form factor, it is sufficient to consider the region $\alpha<0$.

In order to extract the small-$x$ behaviour in the region $\alpha<0$, we split the integral into two regions, $\phi\in(-\infty,0)$ and $\phi\in(0,\infty)$. In each region we pull a factor out of the fraction to leave a denominator of the form $1+\eta$ where $|\eta|\rightarrow0$ as $|\phi|\rightarrow\infty$, and we then Taylor expand each fraction. This gives
\begin{align}
 \langle\vac|\Psi_{R}^{\dagger}(x,0)\sigma_{\alpha}(0)|\theta\rangle_{+}& \sim  c_{\alpha}m^{\alpha^{2}+1/2}\frac{\sin(\pi\alpha)}{\pi}e^{-i\pi\alpha/2}\times \nonumber  \\
 & \left[ \int_{0}^{\infty}{\rm d}\phi\lt(-e^{(\alpha+1/2)\theta-\alpha\phi+xE_{\phi}}-ie^{(\alpha+3/2)\theta-(\alpha+1)\phi+xE_{\phi}}+\ldots\rt)+ \right. \nonumber \\
 & \left. \int_{-\infty}^{0}{\rm d}\phi\lt(-ie^{(\alpha-1/2)\theta-(\alpha-1)\phi+xE_{\phi}}-e^{(\alpha-3/2)\theta-(\alpha-2)\phi+xE_{\phi}}+\ldots\rt)\right]. \label{eq:Rdaggeralpha}
\end{align}
The leading divergent term for $\alpha<0$ occurs in the region $(0,\infty)$. This term can be rewritten using a modified Bessel function:
\begin{align} \label{eq:Besseldiv}
 \int_{0}^{\infty}{\rm d}\phi\,(e^{-\alpha\phi+xE_\phi}) &= \int_{-\infty}^{\infty}{\rm d}\phi\,(e^{-\alpha\phi+mx\cosh\phi})-\int_{-\infty}^{0}{\rm d}\phi\,(e^{-\alpha\phi+mx\cosh\phi}) \nonumber  \\
 &= 2K_{|\alpha|}(-mx)+O(1)
\end{align}
as $x\rightarrow0^-$. Hence, the latter integral diverges like
\begin{equation}
 \Gamma(-\alpha)\left(\frac{2}{-mx}\right)^{-\alpha}.
\end{equation}
The first equation in (\ref{ftw}) allows us to identify the coefficient of the divergence with the one-particle form factor we are seeking. We obtain:
\begin{equation}
 \langle\vac|\sigma_{\alpha+1,\alpha}(0)|\theta\rangle_{+}=c_{\alpha}\frc{e^{-i\pi\alpha/2}}{\Gamma(1+\alpha)}
 m^{\alpha^{2}+\alpha+1/2}e^{(\alpha+1/2)\theta}. \label{eq:defsigma(a+1,a)}
\end{equation}
By analytic continuation, this provides the form factors for all $\alpha\in\R\setminus \Z^*$.

In order to verify the full $z$-dependence of the leading behaviour (\ref{ftw}), we may use the spins of the operators involved and examine the effects of a relativistic boost of rapidity $\beta$ about the origin. Acting with the boost operator $R_{\beta}$, defined essentially by $R_\beta^\dag|\theta\ket = |\theta+\beta\ket$, we see that
\begin{eqnarray}
 R_{\beta}\Psi_{R}^{\dagger}(x,y)R_{\beta}^{\dagger}=e^{\beta/2}\Psi_{R}^{\dagger}(x_\beta,y_\beta) & \text{and} & R_{\beta}\sigma_{\alpha}(0)R_{\beta}^{\dagger}=\sigma_{\alpha}(0)
\end{eqnarray}
as the fields have spins $1/2$ and $0$ respectively. Here, $x_\beta$ and $y_\beta$ are obtained using $z_\beta = e^{\beta}z$ and $\b{z}_\beta = e^{-\beta}z$. Moreover, from the form factor (\ref{eq:defsigma(a+1,a)}), we find that
\beq
	R_{\beta}\sigma_{\alpha+1,\alpha}(0)R_{\beta}^{\dagger}=e^{(\alpha+1/2)\beta}\sigma_{\alpha}(0).
\eeq
Hence, the field $\sigma_{\alpha+1,\alpha}$ has spin $\alpha+1/2$ (as mentioned above). When acting with the rotation operators on $\Psi_{R}^{\dagger}(x,y)\sigma_{\alpha,\alpha}(0)$, we can either perform the boost first and then use the OPE:
\begin{align}
 R_{\beta}\Psi_{R}^{\dagger}(x,y)\sigma_{\alpha}(0)R_{\beta}^{\dagger} & =e^{\beta/2}\Psi_{R}^{\dagger}(x_\beta,y_\beta)\sigma_{\alpha}(0) \nonumber \\
 & \sim e^{\beta/2}(-ize^{\beta})^{\alpha}\sigma_{\alpha+1,\alpha}(0)
\end{align}
or vice versa:
\begin{align}
 R_{\beta}\Psi_{R}^{\dagger}(x,y)\sigma_{\alpha}(0)R_{\beta}^{\dagger} & \sim (-iz)^{\alpha}R_{\beta}\sigma_{\alpha+1,\alpha}(0)R_{\beta}^{\dagger} \nonumber \\
 & = (-iz)^{\alpha}e^{(\alpha+1/2)\beta}\sigma_{\alpha+1,\alpha}(0).
\end{align}
Agreement confirms that the first equation of (\ref{ftw}) is satisfied.

The one particle form factor of the field $\sigma_{\alpha-1,\alpha}(0)$ can be calculated in a similar way, using the second relation of (\ref{ftw}). The calculation, in the range $\alpha>0$, gives
\beq
 \langle\vac|\sigma_{\alpha-1,\alpha}(0)|\theta\rangle_{-} = -ic_{\alpha}\frac{e^{i\pi\alpha/2}}{\Gamma(1-\alpha)}m^{\alpha^{2}-\alpha+1/2}e^{(-\alpha+1/2)\theta} \label{eq:defsigma(a-1,a)},
\eeq
which, by analytic continuation, holds for all $\alpha\in\R\setminus\Z^*$. Note that
\[
	\langle\vac|\sigma_{\alpha-1,\alpha}(0)|\theta\rangle_{-} =
		-i\langle\vac|\sigma_{-\alpha+1,-\alpha}(0)|\theta\rangle_{+}.
\]
With the crossing symmetry relation $\langle\vac|\sigma_{\alpha-1,\alpha}(0)|\theta+i\pi\rangle_{-}=
{}^+\langle\theta|\sigma_{\alpha-1,\alpha}(0)|\vac\rangle$ along with the hermiticity relation (\ref{hermdesc}), this implies
\[
		-i\langle\vac|\sigma_{\alpha-1,\alpha}(0)|\theta+i\pi\rangle_{+}=
		(\langle\vac|\sigma_{\alpha-1,\alpha}(0)|\theta\rangle_{+})^*
\]
which is indeed satisfied by (\ref{eq:defsigma(a+1,a)}).


\subsection{A recursion relation for $c_{\alpha}$, and further verifications of short-distance behaviours} \label{ssrec}

We now proceed to verify the relations (\ref{ftwfull}) and (\ref{consftwfull}), as well as (\ref{ts}) and other similar relations. As we will show, these in fact provide a non-trivial recursion relation for the constant $c_\alpha$. This recursion relation can in fact be used to {\em evaluate} the constant $c_\alpha$, since, for instance, relations (\ref{ftwfull}) were deduced solely from CFT arguments. We note, in particular, that this way of evaluating $c_\alpha$ implements the CFT normalisation of the twist fields somewhat more clearly than in previous methods \cite{LZ96}.

First, we may verify the first relation of (\ref{ftwfull}) in the range $\alpha>1$ by evaluating the leading diverging term of $\bra\vac|\Psi_R^\dag(x)\sigma_\alpha(0)|\theta\ket_+$ for $\alpha$ in that range. Starting with (\ref{eq:Rdaggeralpha}), we need only to consider the region of integration $(-\infty,0)$. We can re-write the integral of the first term displayed using a modified Bessel function, in a similar fashion to what was done in the previous subsection. Comparing with the mass term in the first relation of (\ref{ftwfull}), we find the equation
\begin{equation}
 \langle\vac|\sigma_{\alpha,\alpha-1}(0)|\theta\rangle_{+} = ic_{\alpha}\frac{e^{-i\pi\alpha/2}}{\Gamma(1-\alpha)}m^{\alpha^{2}-\alpha+1/2}e^{\theta(\alpha-1/2)} \label{eq:defsigma(a,a-1)}
\end{equation}
After a shift $\alpha\mapsto\alpha+1$, comparison with the form factor (\ref{eq:defsigma(a+1,a)}) gives the relation
\begin{equation}
 \frac{c_{\alpha}}{c_{\alpha+1}}=\frac{\Gamma(\alpha+1)}{\Gamma(-\alpha)}. \label{eq:rationofc}
\end{equation}
It is a simple matter to verify that this relation is indeed satisfied by (\ref{calpha}) using the properties of the Gamma function and of Barnes' G-function. We remark that this provides a very non-trivial check of the validity of the definition of the fields $\sigma_{\alpha\pm1,\alpha}$ by analytic continuation.

The recursion relation (\ref{eq:rationofc}) along with the symmetry property $c_{\alpha} = c_{-\alpha}$ gives a way of evaluating $c_\alpha$. Indeed, although these relations have many solutions, the solution is made unique by imposing additionally the normalisation condition $c_0=1$ and the conditions of monotonicity and convexity in the range $\alpha\in[0,1)$. These extra conditions are indeed satisfied by (\ref{calpha}). The normalisation condition $c_0=1$ is a simple consequence of the fact that $\sigma_\alpha$ is the identity operator at $\alpha=0$. However, we do not yet have clear arguments for monotonicity and convexity from quantum field theory. We note that all calculations in later sections are unaffected by the explicit choice of the solution $c_\alpha$, and only follow from the recursion relation (\ref{eq:rationofc}) (along with, implicitly, the symmetry property and the normalisation condition).

Second, we may complete the verification of the first relation of (\ref{ftwfull}) by analysing the region $0<\alpha<1$. We first note that the $x=0$ value of the integral in the first line of (\ref{integral}) is
\beq
	\int{\rm d}\phi\,e^{-i\pi/4}e^{\phi/2}\frac{e^{\alpha(\theta-\phi)}}{\sinh(\frac{\theta-\phi}{2}+\frac{i\pi}{4})}
	= -\frc{2\pi e^{i\pi\alpha/2}}{\sin(\pi \alpha)} e^{\theta/2}.
\eeq
This can be obtained by first shifting $\phi\mapsto \phi+\theta$ in order to extract the $\theta$-dependent factor, then by shifting the contour $\phi\mapsto\phi-2i\pi$, getting the pole at $\phi = -3i\pi/2$, and noticing that the integral with a shifted contour is just the initial integral up to a phase (solving the resulting equation gives the answer). Hence, we indeed find that the quantity (\ref{integral}) decays as $x\to0$. The power law of the decay can be obtained simply by taking a derivative with respect to $x$: it is then a divergent behaviour, hence can be obtained by the techniques illustrated above, using (\ref{eq:Rdaggeralpha}). We then integrate over $x$ in order to recover the leading OPE. The leading term is different in the regions $\alpha<1/2$ and $\alpha>1/2$. For $\alpha<1/2$, we recover the first equation of (\ref{ftw}), with (\ref{eq:defsigma(a+1,a)}). For $\alpha>1/2$, taking into account the conditions on the spin, the power law is $(i\b{z})^{1-\alpha}$, and the coefficient is identified with a multiple of the one-particle form factor of the field $m\sigma_{\alpha,\alpha-1}$. The proportionality constant can be evaluated using (\ref{calpha}), and this again only involves verifying the recursion relation (\ref{eq:rationofc}). The resulting leading OPE is in agreement with the first equation of (\ref{ftwfull}) in the range $1/2<\alpha<1$. The case $\alpha=1/2$ simply involves the two terms calculated, again in agreement with (\ref{ftwfull}).

Third, in a similar way, we may verify agreement of the form factor (\ref{eq:defsigma(a-1,a)}) with the second relation of (\ref{ftwfull}) by considering the ranges $\alpha<-1$ and $-1<\alpha<0$, and by using similar arguments as those above. For the range $\alpha<-1$, the mass term of the second relation of (\ref{ftwfull}) leads to the equation
\beq
 \langle\vac|\sigma_{\alpha,\alpha+1}(0)|\theta\rangle_{-} = c_{\alpha}\frac{e^{i\pi\alpha/2}}{\Gamma(1+\alpha)}m^{\alpha^{2}+\alpha+1/2}e^{\theta(-\alpha-1/2)}. \label{eq:defsigma(a,a+1)}
\eeq
Again, comparing with (\ref{eq:defsigma(a-1,a)}), this leads to the recursion relation (\ref{eq:rationofc}), hence is satisfied by (\ref{calpha}). Likewise, the relations (\ref{consftwfull}) can be verified along entirely similar lines.

An alternative way of deriving a recursion relation for the constant $c_{\alpha}$ is to notice that multiple applications of the Fermi fields can lead back to the family of primary twist fields $\sigma_\alpha$. By rearranging (\ref{ts}) we have that
\begin{equation}\label{eqsap}
 \sigma_{\alpha+1}(0)\sim i(-iz)^{-\alpha}{\cal T}[\Psi_{R}^{\dagger}(x,y)\sigma_{\alpha,\alpha+1}(0)]
\end{equation}
as $z\rightarrow0$ (for $\alpha<0$). The vacuum expectation value of $\sigma_{\alpha+1}(0)$ can be calculated from this relation in the same way as the one particle form factors were calculated in Section \ref{sec:1pff}. Firstly (\ref{eq:resofid}) is inserted between the fields to give
\begin{align}
 \langle\vac|\Psi_{R}^{\dagger}(x,y)\sigma_{\alpha,\alpha+1}(0)|\vac\rangle = & \int{\rm d}\theta\, \langle\vac|\Psi_{R}^{\dagger}(x,y)|\theta\rangle_{+}^{\phantom{+}+}\langle\theta|\sigma_{\alpha,\alpha+1}(0)|\vac\rangle \nonumber  \\
 = & \sqrt{m}\int{\rm d}\theta\, e^{\theta/2}e^{-yE_{\theta}+ixp_{\theta}}\langle\vac|\sigma_{\alpha,\alpha+1}(0)|\theta+i\pi\rangle_{-} \nonumber  \\
 = & -ic_{\alpha}m^{\alpha^{2}+\alpha+1}\frac{e^{-i\pi\alpha/2}}{\Gamma(1+\alpha)}\int{\rm d}\theta\, e^{-\alpha\theta-yE_{\theta}+ixp_{\theta}}.
\end{align}
Doing as before, we set $y=0$ and shift the integration contour, $\theta\mapsto\theta-i\pi/2$. We then take the small $x$ expansion to obtain:
\begin{align}
 \langle\vac|\Psi_{R}^{\dagger}(x,0)\sigma_{\alpha,\alpha+1}(0)|\vac\rangle = & -ic_{\alpha}m^{\alpha^{2}+\alpha+1}\frac{1}{\Gamma(1+\alpha)}\int{\rm d}\theta\, e^{-\alpha\theta+xE_{\theta}} \nonumber  \\
 \sim & -ic_{\alpha}m^{\alpha^{2}+\alpha+1}\frac{\Gamma(-\alpha)}{\Gamma(1+\alpha)}\left(\frac{2}{-mx}\right)^{-\alpha},
\end{align}
where on the second line we assumed $\alpha<0$. Hence (\ref{eqsap}) gives
\begin{equation}
 \langle\vac|\sigma_{\alpha+1}(0)|\vac\rangle = c_{\alpha+1}m^{(\alpha+1)^{2}} = c_{\alpha}m^{\alpha^{2}+2\alpha+1}\frac{\Gamma(-\alpha)}{\Gamma(\alpha+1)}
\end{equation}
which gives again the recursion relation (\ref{eq:rationofc}).

Finally, it is worth mentioning that the matrix elements $\bra\vac|\sigma_\alpha(0)\Psi_{R,L}^\dag(x)|\theta\ket_\ep$ and\\ $\bra\vac|\sigma_\alpha(0)\Psi_{R,L}(x)|\theta\ket_\ep$ can be evaluated by similar methods, but without the need for using crossing symmetry. The result is in agreement with the exchange relations: for instance, for $x<0$, the same function of $x$ and $\theta$ is obtained as for the matrix elements $\bra\vac|\Psi_{R,L}^\dag(x)\sigma_\alpha(0)|\theta\ket_\ep$ and $\bra\vac|\Psi_{R,L}(x)\sigma_\alpha(0)|\theta\ket_\ep$ respectively.


\subsection{Higher-particle form factors} \label{sec:3pff}

All higher-particle form factors can be obtained by Wick theorem. The form factor
\[
	\langle\vac|\sigma_{\alpha\pm1,\alpha}(0)|\theta_{1},\ldots,\theta_{2n+1}\rangle_{\epsilon_{1},\ldots,\epsilon_{2n+1}}
\]
is the sum over every term obtained by a product of $n$ (normalised) two-particle form factors of $\sigma_\alpha$ associated to $n$ disjoint pairs of particles (contractions), multiplied by the one-particle form factor of $\sigma_{\alpha\pm1,\alpha}$ associated to the remaining unpaired particle, with an appropriate sign depending on the number of crossings of the contractions:
\beqa
	\lefteqn{
	\langle\vac|\sigma_{\alpha\pm1,\alpha}(0)|\theta_{1},\ldots,\theta_{2n+1}\rangle_{\epsilon_{1},\ldots,\epsilon_{2n+1}} }
	&&\\
	&=& \sum_{{\rm partition}\ \{q\,;\,(s_j,t_j):j=1,\ldots,n\}\atop {\rm of}\ \{1,\ldots,2n-1\}}
		(-1)^{{\rm crossings}} \lt(\prod_{j=1}^n
		\frc{\bra\vac|\sigma_{\alpha}(0)|\theta_{s_j},\theta_{t_j}\ket_{\epsilon_{s_j},\epsilon_{t_j}}}{c_\alpha m^{\alpha^2}}\rt)
		\bra\vac|\sigma_{\alpha\pm1,\alpha}(0)|\theta_q\ket_{\epsilon_q}.\no
\eeqa


\section{The double model} \label{sec:DoubleMod}

In order to make progress in calculating two-point correlation functions of the twist fields we take inspiration from \cite{FZ03} where two non-interacting copies of Ising field theory are considered. Since each copy is quadratic, there is invariance under rotations amongst the two copies, which is an extra continuous symmetry. In the Ising model, the associated conserved charge allowed differential equations for the two-point spin-spin correlation function (and for other objects) to be written down.

We consider two copies of the Dirac model defined in Section \ref{sec:model}, which we denote by $\Psi$ and $\Phi$. These copies have the creation and annihilation operators $D_{\pm}^{(\dagger)}(\theta)$ and $E_{\pm}^{(\dagger)}(\theta)$ respectively, and independently satisfy all the relations of Section \ref{sec:model}. The two copies do not interact, hence obey the anti-commutation relations:
\begin{eqnarray}
 \{\Psi_{R,L}^{(\dagger)}(z),\Phi_{R,L}^{(\dagger)}(w)\}=0 & \text{and} & \{D_{\pm}^{(\dagger)}(\theta_{1}),E_{\pm}^{(\dagger)}(\theta_{2})\}=0.
\end{eqnarray}
There are also independent twist fields associated to the two independent $U(1)$ symmetries of the two copies. These twist fields will, when relevant, carry an extra superscript to denote the copy they belong to: for instance, $\sigma_\alpha^\Psi$ and $\sigma_\alpha^\Phi$. When convenient, we will use such superscripts more generally to denote operators belonging to the respective copies.


\subsection{Conserved charges}

Energy and momentum are conserved charges of the single-copy model. They are associated to the dynamical invariance under space and time translation: the equations of motion (\ref{eom}) possesses this invariance. In the double-copy model, the energy and momentum conserved charges are the sums of the corresponding charges of each copy. However, there are other conserved charges in the double-copy model that can be constructed out of these. As the two copies are non-interacting, the energy-momentum operators for each are still independently conserved quantities, and can be combined differently to give new conserved quantities. The two specific conserved charges that are of interest to us are the differences of those of the single copies. We define $P$ and $\bar{P}$ via the following action (these conserved charges are chosen to be anti-Hermitian):
\begin{align}
 [P,\mathcal{O}^{\Psi}\mathcal{O}^{\Phi}]=& i\partial\mathcal{O}^{\Psi}\mathcal{O}^{\Phi}-i\mathcal{O}^{\Psi}\partial\mathcal{O}^{\Phi} \nonumber \\
 [\bar{P},\mathcal{O}^{\Psi}\mathcal{O}^{\Phi}]=& -i\bar{\partial}\mathcal{O}^{\Psi}\mathcal{O}^{\Phi}+i\mathcal{O}^{\Psi}\bar{\partial}\mathcal{O}^{\Phi}. \label{eq:actionPPbar}
\end{align}
This holds for any local fields $\mathcal{O}^{\Psi}$ and $\mathcal{O}^{\Phi}$ interacting non-trivially with fields in copy $\Psi$ and $\Phi$ respectively. In particular, the actions on the creation and annihilation operators are
\beqa
	&& [P,D_\pm(\theta)] = -ime^\theta D_\pm(\theta),\quad [P,E_\pm(\theta)] = ime^\theta E_\pm(\theta),\n
	&& [\b{P},D_\pm(\theta)] = ime^{-\theta} D_\pm(\theta),\quad [\b{P},E_\pm(\theta)] = -ime^{-\theta} E_\pm(\theta) \no
\eeqa
from which it is simple to derive an explicit expression for $P$ and $\b{P}$ through bilinears in the creation and annihilation operators.

In the double-copy model, there is yet another conserved charge $Z$, related to the $O(2)$ rotation symmetry amongst the copies. Written in terms of the Fermi fields $\Psi$ and $\Phi$, the charge $Z$ is (again, chosen to be anti-Hermitian)
\begin{equation}
 Z=\frac{1}{4\pi}\int{\rm d}x(\Psi_{R}\Phi_{R}^{\dagger}+\Psi_{L}\Phi_{L}^{\dagger}+\Psi_{R}^{\dagger}\Phi_{R}+\Psi_{L}^{\dagger}\Phi_{L}). \label{eq:defZ}
\end{equation}
The action of this charge on the creation and annihilation operators is
\begin{eqnarray}
 [Z,D_{\pm}^{\dagger}(\theta)]=-E_{\pm}^{\dagger}(\theta) & [Z,E_{\pm}^{\dagger}(\theta)]=D_{\pm}^{\dagger}(\theta) \label{eq:actionZ}
\end{eqnarray}
and similarly for $D_{\pm}(\theta)$ and $E_{\pm}(\theta)$.

It is of course possible to obtain higher order conserved charges by commutations amongst $P$, $\b{P}$ and $Z$. In the approach of \cite{FZ03}, where the Ising model was studied, it is assumed that $P$, $\bar{P}$ and $Z$ all preserve the quantum state where correlation functions are evaluated, and the Ward identities corresponding to certain of these higher order conserved charges are used to derive differential equations for twist-field correlation functions (it was also observed there that the higher order conserved charges form a $\widehat{sl(2)}$ algebra). However, it turns out that this is not necessary, as was noticed in \cite{Dth}. Indeed, the Ward identities coming from the invariance under the $Z$-action alone are sufficient to obtain non-trivial equations for correlation functions -- we do not need to assume that the dynamical space-time translation symmetry is a symmetry of the quantum state under study. When the quantum state does possess this symmetry, the space-time dependence of correlation functions simplifies, and the resulting equations are the integrable differential equations describing the Ising correlation functions. As we will see, the same structure occurs in the present case of the Dirac theory (with the additional condition that we need parity invariance in order to further simplify the space-time dependence of correlation functions). This is an important remark, as it shows that the method may have applicability beyond the cases where the quantum states are invariant under space-time translation (and parity).

Below, we will derive the integrable differential equations from the Ward identities associated to $Z$, and use the simpler space-time dependence of correlation functions coming from translation and parity symmetries. We refer to Appendix \ref{sec:eq-rtp} for the equations resulting from the $Z$ Ward identities when these symmetries are not present.

As was observed in \cite{Dth} (in the generalised situation of a theory on the Poinar\'e disk), it is the relation
\begin{align}\label{eqmass}
 [P,[\bar{P},Z]]=[\bar{P},[P,Z]]=4m^{2}Z
\end{align}
which provides the dependence on the mass in the differential equations. Relation (\ref{eqmass}) corresponds essentially to the equations of motion of the theory, as expressed using the charge $Z$. It can be derived from (\ref{eq:actionPPbar}) and (\ref{eq:defZ}) using the equations of motion (\ref{eom}). Relation (\ref{eqmass}) is the only one where the massive theory is used: besides it, we only need the action of $Z$ on products of twist fields and their first derivatives in order to find the differential equations, and this action, as is verified below, is solely a consequence of CFT calculations (i.e. calculations using the massless limit of the theory).


\subsection{Equations for the action of $Z$}\label{sec:eqZ}

The action of the conserved charge $Z$ on products of local fields $\mathcal{O}^{\Psi}\mathcal{O}^{\Phi}$ gives linear combinations of similar products of local fields (with unchanged locality index) whenever the combined locality index of the product $\mathcal{O}^{\Psi}\mathcal{O}^{\Phi}$ is one. Otherwise, the result of the action is non-local, and the resulting Ward identities are not useful. We expect that the relations derived in the CFT limit in Appendix \ref{sec:cft}, where the right- and left-moving components of the Fermions are separated, will carry across to the massive case when written in terms of well-defined twist fields of the massive theory. Although we have not provided a full analysis of this fact, which would involve generalising the OPEs derived in Appendix \ref{sec:cft} to massive-theory OPEs involving mass terms as in (\ref{ftwfull}) and (\ref{consftwfull}), we have nevertheless verified it using form factors with up to 4 particles.

The equations representing the action of the charge $Z$ on twist fields involving no derivatives are:
\begin{subequations} \label{eq:1stzrels}
\beqa
 [Z,\sigma_{\alpha}^{\Psi}\sigma_{\alpha}^{\Phi}] &= & 0  \label{eq:[Z,aa]} \\
{} [Z,\sigma_{\alpha}^{\Psi}\sigma_{\alpha-1}^{\Phi}] &= & i(\sigma_{\alpha-1,\alpha}^{\Psi}\sigma_{\alpha,\alpha-1}^{\Phi}-\sigma_{\alpha,\alpha-1}^{\Psi}\sigma_{\alpha-1,\alpha}^{\Phi}) \label{eq:[Z,aa-1]} \\
{} [Z,\sigma_{\alpha}^{\Psi}\sigma_{\alpha+1}^{\Phi}] &= & i(\sigma_{\alpha+1,\alpha}^{\Psi}\sigma_{\alpha,\alpha+1}^{\Phi}-\sigma_{\alpha,\alpha+1}^{\Psi}\sigma_{\alpha+1,\alpha}^{\Phi}) \label{eq:[Z,aa+1]} \\
{} [Z,\sigma_{\alpha+1,\alpha}^{\Psi}\sigma_{\alpha,\alpha+1}^{\Phi}] &= & i( \sigma_{\alpha}^{\Psi}\sigma_{\alpha+1}^{\Phi}-\sigma_{\alpha+1}^{\Psi}\sigma_{\alpha}^{\Phi}) \label{eq:[Z,mix]} \\
{} [Z,\sigma_{\alpha+1}^{\Psi}\sigma_{\alpha-1}^{\Phi}] &=&\frac{1}{\alpha}\left((\partial\sigma_{\alpha,\alpha+1}^{\Psi})\sigma_{\alpha,\alpha-1}^{\Phi}-\sigma_{\alpha,\alpha+1}^{\Psi}(\partial\sigma_{\alpha,\alpha-1}^{\Phi})+\rt.\n &&
\lt.\quad\ (\bar{\partial}\sigma_{\alpha+1,\alpha}^{\Psi})\sigma_{\alpha-1,\alpha}^{\Phi}-\sigma_{\alpha+1,\alpha}^{\Psi}(\bar{\partial}\sigma_{\alpha-1,\alpha}^{\Phi})\right)  \label{eq:[Z,a+1a-1]} \\
{} [Z,\sigma_{\alpha-1}^{\Psi}\sigma_{\alpha+1}^{\Phi}]&=&\frac{1}{\alpha}\left(\sigma_{\alpha,\alpha-1}^{\Psi}
(\partial\sigma_{\alpha,\alpha+1}^{\Phi})-(\partial\sigma_{\alpha,\alpha-1}^{\Psi})\sigma_{\alpha,\alpha+1}^{\Phi}+ \rt.\n &&
\lt.\qquad\ \sigma_{\alpha-1,\alpha}^{\Psi}(\bar{\partial}\sigma_{\alpha+1,\alpha}^{\Phi})-(\bar{\partial}\sigma_{\alpha-1,
\alpha}^{\Psi})\sigma_{\alpha+1,\alpha}^{\Phi}\right). \label{eq:[Z,a-1a+1]} 
\eeqa
\end{subequations}
Here and in the rest of this section, all fields are evaluated at one single point, say for instance the point $(0,0)$. When verifying these equations we place the vacuum state on the left and a multi-particle state on the right. The left-hand side is expanded using (\ref{eq:actionZ}) along with $\langle\vac|Z=0$ and $Z|\vac\ket=0$, this gives form factors of the type (\ref{eq:Formfactorsaa}). The right-hand side gives form factors of the type found in Sections \ref{sec:1pff} and \ref{sec:3pff}. The two sides are found to be equal. We note that for this equality to occur, we do not need explicitly the form (\ref{calpha}) of the vacuum expectation values, but only the recursion relation (\ref{eq:rationofc}). This agreement is very non-trivial and is strong evidence that Equations (\ref{eq:1stzrels}) hold for all matrix elements.

The origins of Equations (\ref{eq:1stzrels}) are not obvious without reference to the CFT. Equation (\ref{eq:[Z,aa]}), however, is easy to verify, and using this (\ref{eq:[Z,aa-1]}) and (\ref{eq:[Z,aa+1]}) can be `derived' from the relations of the type
\begin{eqnarray*}
 \Psi_{R}^{\dagger}\Psi_{L}\sigma_{\alpha}\mapsto \sigma_{\alpha+1} & \text{and} & \Psi_{R}\Psi_{L}^{\dagger}\sigma_{\alpha}\mapsto \sigma_{\alpha-1}
\end{eqnarray*}
(see (\ref{consftw}) and (\ref{ts}) for the more accurate relations). The final three equations do not appear to have simple origins outside of the CFT but can be verified using the method described above. It is also worth noting that the left-hand sides of (\ref{eq:[Z,aa-1]}) and (\ref{eq:[Z,aa+1]}) are the same up to a shift in indices and a relabelling of the fields. The right-hand sides are also consistent with these operations.

Equations involving one derivative are as follows, and have been derived from CFT and verified using form factors like those above:
\begin{eqnarray*}
 [Z,(\partial\sigma_{\alpha}^{\Psi})\sigma_{\alpha}^{\Phi}] &= & \alpha(\sigma_{\alpha-1,\alpha}^{\Psi}\sigma_{\alpha+1,\alpha}^{\Phi}-\sigma_{\alpha+1,\alpha}^{\Psi}\sigma_{\alpha-1,\alpha}^{\Phi})  \\
{} [Z,(\bar{\partial}\sigma_{\alpha}^{\Psi})\sigma_{\alpha}^{\Phi}] &= & \alpha(\sigma_{\alpha,\alpha-1}^{\Psi}\sigma_{\alpha,\alpha+1}^{\Phi}-\sigma_{\alpha,\alpha+1}^{\Psi}\sigma_{\alpha,\alpha-1}^{\Phi})  \\
{} [Z,(\partial\sigma_{\alpha}^{\Psi})\sigma_{\alpha+1}^{\Phi}] &= & i((\partial\sigma_{\alpha,\alpha+1}^{\Psi})\sigma_{\alpha+1,\alpha}^{\Phi}-\sigma_{\alpha+1,\alpha}^{\Psi}(\partial\sigma_{\alpha,\alpha+1}^{\Phi}))  \\
{} [Z,(\bar{\partial}\sigma_{\alpha}^{\Psi})\sigma_{\alpha+1}^{\Phi}] &= & i(\sigma_{\alpha,\alpha+1}^{\Psi}(\bar{\partial}\sigma_{\alpha+1,\alpha}^{\Phi})-(\bar{\partial}\sigma_{\alpha+1,\alpha}^{\Psi})\sigma_{\alpha,\alpha+1}^{\Phi})  \\
{} [Z,(\partial\sigma_{\alpha}^{\Psi})\sigma_{\alpha-1}^{\Phi}] &= & i((\partial\sigma_{\alpha,\alpha-1}^{\Psi})\sigma_{\alpha-1,\alpha}^{\Phi}-\sigma_{\alpha-1,\alpha}^{\Psi}(\partial\sigma_{\alpha,\alpha-1}^{\Phi}))  \\
{} [Z,(\bar{\partial}\sigma_{\alpha}^{\Psi})\sigma_{\alpha-1}^{\Phi}] &= & i(\sigma_{\alpha,\alpha-1}^{\Psi}(\bar{\partial}\sigma_{\alpha-1,\alpha}^{\Phi})-(\bar{\partial}\sigma_{\alpha-1,\alpha}^{\Psi})\sigma_{\alpha,\alpha-1}^{\Phi})  \\
{} [Z,(\partial\sigma_{\alpha,\alpha+1}^{\Psi})\sigma_{\alpha+1,\alpha}^{\Phi}] &= & i(\sigma_{\alpha+1}^{\Psi}(\partial\sigma_{\alpha}^{\Phi})-(\partial\sigma_{\alpha}^{\Psi})\sigma_{\alpha+1}^{\Phi})  \\
{} [Z,(\partial\sigma_{\alpha+1,\alpha}^{\Psi})\sigma_{\alpha,\alpha+1}^{\Phi}]&= & i(\sigma_{\alpha}^{\Psi}(\partial\sigma_{\alpha+1}^{\Phi})-(\partial\sigma_{\alpha+1}^{\Psi})\sigma_{\alpha}^{\Phi})  \\
{} [Z,(\bar{\partial}\sigma_{\alpha,\alpha+1}^{\Psi})\sigma_{\alpha+1,\alpha}^{\Phi}]&= & i((\bar{\partial}\sigma_{\alpha+1}^{\Psi})\sigma_{\alpha}^{\Phi}-\sigma_{\alpha}^{\Psi}(\bar{\partial}\sigma_{\alpha+1}^{\Phi})) \\
{} [Z,(\bar{\partial}\sigma_{\alpha+1,\alpha}^{\Psi})\sigma_{\alpha,\alpha+1}^{\Phi}]&= & i((\bar{\partial}\sigma_{\alpha}^{\Psi})\sigma_{\alpha+1}^{\Phi}-\sigma_{\alpha+1}^{\Psi}(\bar{\partial}\sigma_{\alpha}^{\Phi})).
\end{eqnarray*}
Note that in Section \ref{sec:corfun}, we only need few of the zero- and one-derivative equations showed here.

Finally, the action of $Z$ on fields involving higher numbers of derivatives are not in general expected, in the massive theory, to be in agreement with calculations from CFT. However, the only ones that we actually need can be obtained by using those above along with the equation of motion (\ref{eqmass}) of the charge $Z$. Indeed, we only need actions of $Z$ on double-derivative fields of the form $[P,[\bar{P},\mathcal{O}^{\Psi}\mathcal{O}^{\Phi}]]$, which we can evaluate using
\begin{align*}
	\lefteqn{[Z,[P,[\bar{P},\mathcal{O}^{\Psi}\mathcal{O}^{\Phi}]]]} & \\ & =
	[[P,[\bar{P},Z]],\mathcal{O}^{\Psi}\mathcal{O}^{\Phi}]+
	[\bar{P},[Z,[P,\mathcal{O}^{\Psi}\mathcal{O}^{\Phi}]]]+
	[P,[Z,[\bar{P},\mathcal{O}^{\Psi}\mathcal{O}^{\Phi}]]]-
	[P,[\bar{P},[Z,\mathcal{O}^{\Psi}\mathcal{O}^{\Phi}]]] \\ &=
	4m^2 [Z,\mathcal{O}^{\Psi}\mathcal{O}^{\Phi}]+
	[\bar{P},[Z,[P,\mathcal{O}^{\Psi}\mathcal{O}^{\Phi}]]]+
	[P,[Z,[\bar{P},\mathcal{O}^{\Psi}\mathcal{O}^{\Phi}]]]-
	[P,[\bar{P},[Z,\mathcal{O}^{\Psi}\mathcal{O}^{\Phi}]]]
\end{align*}
This is how the mass dependence will appear in the equations for correlation functions.


\section{Correlation functions} \label{sec:corfun}

\subsection{Correlation functions of interest}

The aim of this section is to derive differential equations for two-point correlation functions of twist fields using Ward identities associated to the conservation of the charge $Z$. We will keep the state where the correlation functions is evaluated as general as possible.

The first general form of a state invariant under $Z$ is the mixed state where the average of an operator $A$ is given by
\beq\label{av1}
	\frc{\Tr\lt(e^{V} A\rt)}{\Tr\lt(e^V\rt)},
\eeq
where $V$ is any bilinear form in the creation and annihilation operators (or in the Dirac Fermi fields). In the double copy model, it is clear that $V^\Psi+V^\Phi$ is invariant under $O(2)$ rotations, hence that
\beq\label{av2}
	\frc{\Tr\lt(e^{V^\Psi + V^\Phi} A\rt)}{\Tr\lt(e^{V^\Psi + V^\Phi}\rt)}
\eeq
is zero whenever $A$ is of the form $[Z,\cdots]$: these are the Ward identities associated to $Z$. These states include, for instance, the finite-temperature state $V = -H/T$, with temperature $T$, as well as the vacuum state, the limit $T\to0$.

Another form of the quantum state is that where averages are given by
\beq\label{av3}
	\bra B'|A|B\ket
\eeq
where $|B\ket$ and $|B'\ket$ are boundary states of the form $|B\ket = e^V|\vac\ket$ and $|B'\ket = e^{V'}|\vac\ket$, for bilinear forms $V$ and $V'$ in the creation and annihilation operators. Again,
\beq\label{av4}
	\bra (B')^\Psi (B')^\Phi |A|B^\Psi B^\Phi\ket
\eeq
is zero whenever $A$ is of the form $[Z,\cdots]$. Such states include, for instance, the integrable boundary states \cite{GZ94} of the Dirac theory.

Further, since the action of the $Z$ operator on local fields is a local property, it is not affected by the exact space where the quantisation holds: we may also choose a Hilbert space of quantisation on a finite segment (e.g. the circle of length $L$), and again consider the states mentioned above.

We will use the single notation $\bra A\ket$ in order to represent averages (\ref{av1}), (\ref{av2}), (\ref{av3}) or (\ref{av4}).

We will show that the sinh-Gordon partial differential equation determines the following three families of correlation functions:
\beqa
	F(x,y) & = & \bra\sigma_\alpha(x,y) \sigma_\alpha(0)\ket, \nonumber \\
	G(x,y) & = & \bra\sigma_\alpha(x,y) \sigma_{\alpha+1}(0)\ket, \nonumber \\
	H(x,y) & = & \bra\sigma_{\alpha+1,\alpha}(x,y) \sigma_{\alpha,\alpha+1}(0)\ket \\
\eeqa
whenever the correlation functions are translation and parity invariant. Here, the dependence on $\alpha$ is implicit, in order not to clutter the equations. In the same spirit, we will also use the notation
\beq
	\t{F}(x,y) = \bra\sigma_{\alpha+1}(x,y) \sigma_{\alpha+1}(0))\ket.
\eeq
When there is no translation or parity symmetry, more general equations can be derived, see Appendix \ref{sec:eq-rtp}.

We note that correlation functions involving different ordering of the two fields present in the definitions of $F(x,y)$, $G(x,y)$ and $H(x,y)$ are simply related to those written above. This is obvious for the cases of $F(x,y)$ and $G(x,y)$, since the twist fields involved commute. For the case of $H(x,y)$, we must apply the braiding relations (\ref{eq:equaltimebraiding}). Recall that by definition
\beqa
 \sigma_{\alpha+1,\alpha}(x,y)&=&\lim_{z\rightarrow w}(-i(z-w))^{-\alpha}\Psi_{R}^{\dagger}(x',y')\sigma_{\alpha}(x,y) \n
 \sigma_{\alpha,\alpha+1}(x,y)&=&\lim_{z\rightarrow w}(i(\bar{z}-\bar{w}))^{-\alpha}\Psi_{L}(x',y')\sigma_{\alpha}(x,y) 
 \label{defslimit}
\eeqa
where $z=-\frac{i}{2}(x'+iy')$ and $w=-\frac{i}{2}(x+iy)$ are appropriately time ordered ($y'>y$). The factors $(-i(z-w))^{-\alpha}$ and $(i(\bar{z}-\bar{w}))^{-\alpha}$ are taken on the principal branch, and so are continuous exactly where $\Psi_{R}^{\dagger}(x',y')\sigma_{\alpha}(x,y)$ and $\Psi_{L}(x',y')\sigma_{\alpha}(x,y)$ are continuous. Using these definitions and (\ref{eq:equaltimebraiding}) we obtain
\begin{align}\label{braidingdesc}
	\sigma_{\beta,\beta+1}(x_{2},y_{2})\sigma_{\alpha+1,\alpha}(x_{1},y_{1})=\left\{
	\begin{array}{cc}
	-e^{2\pi i\beta} & {\rm if }~ x_{1}>x_{2}  \\
	-e^{2\pi i\alpha} & {\rm if }~ x_{1}<x_{2}
	\end{array} \right\}
	\sigma_{\alpha+1,\alpha}(x_{1},y_{1})\sigma_{\beta,\beta+1}(x_{2},y_{2}).
\end{align}

\subsection{Translation and parity symmetry}\label{sec:syms}

Space and time translation invariance simplifies the functional form of the correlation functions. It implies that it is indeed sufficient to fix the position of the second field to $0$ in all correlation functions above, as we have done, and further that
\beq
	F(-x,-y) = F(x,y),\quad \t{F}(-x,-y) = \t{F}(x,y).
\eeq
It also implies, using the fact that the fields $\sigma_\alpha$ and $\sigma_{\alpha+1}$ commute and using the braiding (\ref{braidingdesc}), that
\beq\label{somesym}
	\bra \sigma_{\alpha+1}(x,y) \sigma_\alpha(0)\ket = G(-x,-y),\quad
	\bra \sigma_{\alpha,\alpha+1}(x,y) \sigma_{\alpha+1,\alpha}(0) \ket = -e^{2\pi i \alpha} H(-x,-y).
\eeq

Space and time parity symmetry (i.e. invariance under simultaneous inversion of space and Euclidean time) further simplifies the functions $G$ and $H$. In two dimensions, this is equivalent to Euclidean rotation by an angle of $\pi$. Since this is a Euclidean rotation, some subtleties occur. The symmetry cannot be expressed simply through a unitary operator, but rather is a relation connecting vacuum expectation values. It says that there exists a unitary operator $R$, which performs a parity symmetry on individual local fields, such that the following equality holds:
\beq\label{parinv}
	\bra\vac|\Or_1(x_1,y_1) \cdots \Or_n(x_n,y_n) |\vac\ket = \bra \vac|R \Or_n(x_n,y_n) \cdots \Or_1(x_1,y_1) R^\dag|\vac\ket
\eeq
whenever the local fields $\Or_j(x_j,y_j)$ are time-ordered: $y_1>\ldots>y_n$. The unitary operator does not act trivially on the vacuum, because it transforms an in-vacuum into an out-vacuum and vice versa. But it acts locally on local fields, in particular inverting both space and time coordinates, so that the correlation function on the right-hand side is still time-ordered. Relation (\ref{parinv}) has its origin in the Euclidean path-integral picture, whose connection to the operatorial picture requires time-ordering: it represents the invariance of Euclidean correlation functions under a rotation of the Euclidean plane by $\pi$.

The operator $R$ can be written explicitly. Naturally, it could be seen as a shift of rapidity by the imaginary number $i\pi$. However, there is another, more appropriate way of representing this operator, with the same effect on Fermion fields: the operator $R$ can be seen as a combination of two symmetries, space parity $S$ and time parity $T$, with
\[
	R=TS.
\]
Space parity is implemented by the unitary linear operator $S$ satisfying
\beq
	S\Psi_{R,L}(x,y)S = \pm i \Psi_{L,R}(-x,y),\quad S^2 = 1,
\eeq
and time parity is implemented by the unitary linear operator $T$ satisfying
\beq
	T\Psi_{R,L}(x,y)T = \Psi_{L,R}(x,-y),\quad T^2 = 1
\eeq
(note that the latter is not an anti-linear operator, because it inverts the Euclidean time instead of the real time -- it is akin to a parity operator). It is easy to verify that these are indeed dynamical symmetries: they preserve the equations of motion (\ref{eom}). The actions of $S$ and $T$ on the mode operators $D_\pm(\theta)$ can be obtained from the relations above, and are given by $SD_\pm(\theta)S = \mp D_\pm(-\theta)$ and $TD_\pm(\theta)T = - iD_\mp^\dag(-\theta)$. The combined parity transformation is
\beq\label{transmodepar}
	TS\Psi_{R,L}(x,y)ST = \pm i \Psi_{R,L}(-x,-y),\quad TSD_\pm(\theta)ST = \pm i D_\mp^\dag(\theta),
\eeq
and it has the following simple operator representation:
\beq
	TS = \exp \lt[\frc{i\pi}2 \int {\rm d}\theta \,
		(D_-^\dag(\theta) D_+^\dag(\theta) + D_+(\theta) D_-(\theta)) \rt].
\eeq

The transformation properties (\ref{transmodepar}) tell us that the creation operators are transformed into annihilation operators of opposite charges, and vice versa. In combination with (\ref{parinv}), this allows us to extend the invariance property to correlation functions in arbitrary states: we have in general, again with time ordering $y_1>\ldots>y_n$,
\beq\label{parinvarb}
	\bra B'|\Or_1(x_1,y_1) \cdots \Or_n(x_n,y_n) |B\ket = \bra \b{B}|R \Or_n(x_n,y_n) \cdots \Or_1(x_1,y_1) R^\dag|\b{B}'\ket
\eeq
where $|\b{B}\ket$ and $|\b{B}'\ket$ are the states obtained from $|B\ket$ and $|B'\ket$, respectively, by inverting the charges of all particles, with extra factors $-i$ and $+i$ for each particle if the new charge is $-$ or $+$ respectively. The action of a rotation by $\pi$ on multi-particle states is indeed clear in the Euclidean picture: for instance, particles coming from the past are rotated to come from the future, and can be re-interpreted as going towards the future if their charge is conjugated (a particle moving forward in time is the same as its anti-particle moving backward in time). In general, the parity invariance property (\ref{parinvarb}) can be generalised to mixed states using the trace definition and (\ref{parinvarb}) on the summand in the trace.

Finally, we may also obtain the transformation property of twist fields. Since $\sigma_\alpha$ is spinless, and since $\sigma_{\alpha+1,\alpha}$ and $\sigma_{\alpha,\alpha+1}$ have opposite spins, it is clear that $\sigma_\alpha(x,y) \sigma_{\alpha+1}(0)$ and $\sigma_{\alpha+1,\alpha}(x,y) \sigma_{\alpha,\alpha+1}(0)$ transform under rotation by rotating the coordinates, without phase factors associated to the spin. There is an extra subtlety, however, having to do with the fact that we only ask for invariance of the state under rotation by the unique angle $\pi$. As a consequence, the correlation functions may depend on the direction of the cut emanating from the twist field, which we have chosen to be towards the right according to (\ref{eq:equaltimebraiding}). After a $\pi$ rotation, the cut goes in the opposite direction. This can be represented by the presence of an extra $U(1)$ transformation operator, so that we have
\beq
	R\sigma_\alpha(x,y) R^\dag = \sigma_\alpha(-x,-y)e^{2\pi i \alpha Q}
\eeq
where $Q$ is the Hermitian $U(1)$ charge. The operator $e^{2\pi i \alpha Q}$ is effectively a branch cut along the whole $x$-axis, cancelling out the twist fields cut to the right and reinstating the cut to the left. From (\ref{defslimit}) (and remembering that relation (\ref{parinv}) inverts the positions of operators), this leads to
\beqa
	R\sigma_{\alpha+1,\alpha}(x,y)R^\dag &=&
		e^{-i\pi (\alpha+1/2)} \sigma_{\alpha+1,\alpha}(-x,-y) e^{2\pi i \alpha Q}\\
	R\sigma_{\alpha,\alpha+1}(x,y)R^\dag &=&
		e^{i\pi (\alpha-1/2)} \sigma_{\alpha,\alpha+1}(-x,-y) e^{2\pi i \alpha Q}.
\eeqa

Hence, a parity transformation gives us, using translation invariance of the state and the last relation (\ref{somesym}),
\beqa
	H(x,y) &=& \bra \sigma_{\alpha+1,\alpha}(x,y) \sigma_{\alpha,\alpha+1}(0)\ket \n
	&=& -\bra \sigma_{\alpha,\alpha+1}(0) e^{2\pi i \alpha Q}
		\sigma_{\alpha+1,\alpha}(-x,-y) e^{2\pi i \alpha Q}\ket^c \n
	&=& -e^{-2 i\pi \alpha} \bra \sigma_{\alpha,\alpha+1}(x,y)
		\sigma_{\alpha+1,\alpha}(0,0) e^{4\pi i \alpha Q}\ket^c \no
\eeqa
where the superscript $c$ means that the state is as in the right-hand side of (\ref{parinvarb}) (or its generalisation to mixed states). Likewise,
\beqa
	G(x,y) &=& \bra\sigma_{\alpha}(x,y)\sigma_{\alpha+1}(0) \ket \n
		&=& \bra\sigma_{\alpha+1}(0)\sigma_\alpha(-x,-y)e^{4\pi i\alpha Q}\ket^c\n
		&=& \bra\sigma_{\alpha+1}(x,y)\sigma_\alpha(0)e^{4\pi i\alpha Q}\ket^c.
\eeqa
Hence, parity invariance for these twist field correlation functions requires the following condition on the state:
\beq\label{condparinv}
	\bra \cdots e^{4\pi i \alpha Q}\ket^c = \bra \cdots \ket.
\eeq
If this condition holds, then we find:
\beq
	G(-x,-y) = G(x,y),\quad H(-x,-y) = H(x,y).
\eeq

An example of a situation where there is translation and parity invariance is the finite-temperature state $V = -H/T$, with additionally the condition that $2\alpha$ be an integer. A correlation function in a finite-temperature state has the geometric meaning of a path integral evaluated on an infinitely long cylinder. The cuts of the twist fields extend towards the ends of the cylinder, and as such, affect the state representing the asymptotic values of the fields there (these asymptotic values are asymptotic states in the quantisation on the circle). This means that correlation functions with cuts extending towards the right and the left are in general different. Hence, indeed parity symmetry imposes that there be no cut extending towards the right or the left. This imposes that the locality factor of the combined fields of the correlation function be one, hence that $2\alpha$ be an integer. Note that by the relation between the Dirac theory and the Ising model, this gives the spin and disorder fields and their descendants. More generally, we see that a correlation function of the form $\bra\sigma_{\alpha}\sigma_{\beta}\ket$, in a finite-temperature state and with the condition $\alpha+\beta\in \Z$, is both translation and parity invariant.


\subsection{Ward identities} \label{sec:wardident}

Before embarking on our analysis of the Ward identities we first note that any correlation function of the form $\langle\sigma_{\alpha,\alpha\pm1}(x,y)\sigma_{\beta,\beta}(0)\rangle$, $\langle\sigma_{\alpha,\alpha+1}(x,y)\sigma_{\alpha,\alpha+1}(0)\rangle$ or $\langle\sigma_{\alpha,\alpha-1}(x,y)\sigma_{\alpha,\alpha-1}(0)\rangle$ vanishes, as a consequence of $U(1)$ charge conservation.

In order to derive a set of differential equations for the functions $F(x,y)$, $\tilde{F}(x,y)$, $G(x,y)$ and $H(x,y)$ we start with the following Ward identities,
\begin{subequations} \label{eq:WardIdents}
 \begin{align}
  \langle[Z,\sigma_{\alpha}^{\Psi}(x,y)\sigma_{\alpha+1}^{\Phi}(x,y)\sigma_{\alpha+1,\alpha}^{\Psi}(0)\sigma_{\alpha,\alpha+1}^{\Phi}(0)]\rangle=0  \\
  \langle[Z,[P,\sigma_{\alpha}^{\Psi}(x,y)\sigma_{\alpha+1}^{\Phi}(x,y)]\sigma_{\alpha+1,\alpha}^{\Psi}(0)\sigma_{\alpha,\alpha+1}^{\Phi}(0)]\rangle=0  \\
  \langle[Z,[\bar{P},\sigma_{\alpha}^{\Psi}(x,y)\sigma_{\alpha+1}^{\Phi}(x,y)]\sigma_{\alpha+1,\alpha}^{\Psi}(0)\sigma_{\alpha,\alpha+1}^{\Phi}(0)]\rangle=0 \\
  \langle[Z,[P,[\bar{P},\sigma_{\alpha}^{\Psi}(x,y)\sigma_{\alpha+1}^{\Phi}(x,y)]]\sigma_{\alpha,\alpha+1}^{\Psi}(0)\sigma_{\alpha+1,\alpha}^{\Phi}(0)]\rangle=0  \\
  \langle[Z,[P,\sigma_{\alpha}^{\Psi}(x,y)\sigma_{\alpha+1}^{\Phi}(x,y)][\bar{P},\sigma_{\alpha,\alpha+1}^{\Psi}(0)\sigma_{\alpha+1,\alpha}^{\Phi}(0)]]\rangle=0
 \end{align}
\end{subequations}
and expand them using the action of $Z$ obtained in Section \ref{sec:eqZ}.

To ease our notation it will be useful to introduce the differential operator
\beq
\mathcal{D}(f,g):=\frac{1}{2}\left(f\partial\bar{\partial}g+g\partial\bar{\partial}f-\partial f\bar{\partial}g-\partial g\bar{\partial}f \right)
\eeq
and also write $\mathcal{D}(f):=\mathcal{D}(f,f)$. We note that
\begin{equation}
 \mathcal{D}(e^{f})=e^{2f}\partial\bar{\partial}f.
\end{equation}

From the Ward identities (\ref{eq:WardIdents}) we obtain, respectively, the equations
\begin{subequations}
 \beqa 
  e^{2\pi i\alpha}H^{2}+G^{2}-F\tilde{F} &=&0 \label{eq:ward1} \\
  F\partial \tilde{F}-(\partial F)\tilde{F} &=&0 \label{eq:ward2} \\
 (\bar{\partial} F)\tilde{F}-F\bar{\partial}\tilde{F} &=&0 \label{eq:ward3} \\
  e^{2\pi i\alpha}\left(\mathcal{D}(H)-2m^{2}H^{2}\right)-\mathcal{D}(G)+\mathcal{D}(F,\tilde{F}) &=&0 \label{eq:ward4} \\
 e^{2\pi i\alpha}\mathcal{D}(H)-\mathcal{D}(G)-\mathcal{D}(F,\tilde{F}) &=&0. \label{eq:ward6}
 \eeqa
\end{subequations}
These look similar in structure to equations derived in \cite{FZ03} for the Ising spin-spin correlator except that here we have 4 functions to solve for.

As $F$ and $\tilde{F}$ are only functions of $x$ and $y$, Equations (\ref{eq:ward2}) and (\ref{eq:ward3}) can be solved to give
\begin{equation} \label{eq:defk}
 \tilde{F}=k^{2}F
\end{equation}
where $k$ is the constant of integration. This relation is somewhat surprising as it tells us the two correlation functions behave in the same way, at both short and long distances, despite the fields having different dimensions.  Using the form factors for the twist fields $k$ can be calculated exactly. Inserting (\ref{eq:resofid}) between the fields in the correlation functions allows us to compare terms involving equal numbers of particles. Clearly the zero particle terms differ by the constant $k^{2}$ where
\begin{equation}
 k=\frac{c_{\alpha+1}}{c_{\alpha}}m^{2\alpha+1}
\end{equation}
and it is a simple exercise to show that the same is true for all subsequent terms. We do not have yet a convincing intuitive explanation of the relation (\ref{eq:defk}); it simply follows from our analytic-continuation definition of descendent twist fields.


\subsection{Differential equations for correlation functions}
Our next step towards an equation for the correlation functions is to define the functions $\chi$ and $\varphi$ via
\begin{eqnarray}
 kF+G & = & e^{\chi}\cosh(\varphi) \n
 kF-G & = & e^{\chi}\sinh(\varphi)
\end{eqnarray}
so that (\ref{eq:ward1})becomes
\begin{equation}
 e^{2\pi i\alpha}H^{2}=\frac{1}{2}e^{2\chi}\sinh(2\varphi).
\end{equation}

It is now possible to write (\ref{eq:ward4}) and (\ref{eq:ward6}) in terms of $\chi$ and $\varphi$ only:
\begin{subequations}
 \begin{eqnarray}
  \sinh(2\varphi)\left(\partial\bar{\partial}\chi+\partial\varphi\bar{\partial}\varphi\right)+\cosh(2\varphi)\left(\partial\bar{\partial}\varphi-\coth(2\varphi)\partial\varphi\bar{\partial}\varphi\right)-m^{2}\sinh(2\varphi) & = & 0 \label{eq:newward5.2} \\
  \sinh(2\varphi)\left(\partial\bar{\partial}\chi+2\partial\varphi\bar{\partial}\varphi-\partial\bar{\partial}\varphi\right)+\cosh(2\varphi)\left(\partial\bar{\partial}\varphi-2\coth(2\varphi)\partial\varphi\bar{\partial}\varphi-\partial\bar{\partial}\chi\right) & = & 0 \label{eq:newward6.2}
 \end{eqnarray}
\end{subequations}
and we can use (\ref{eq:newward5.2}) to eliminate $\chi$ from (\ref{eq:newward6.2}), leaving
\begin{equation}
 \partial\bar{\partial}\varphi-(1+\coth(2\varphi))\partial\varphi\bar{\partial}\varphi+m^{2}\sinh^{2}(2\varphi)(1-\coth(2\varphi))=0. \label{eq:phi1}
\end{equation}

In order to retrieve (\ref{eq:result1}) from (\ref{eq:phi1}) it is necessary to eliminate the $\partial\phi\bar{\partial}\phi$ term. To do this we note that
\begin{equation}
 \frac{f''(\varphi)}{f'(\varphi)}=1+\coth(2\varphi)
\end{equation}
admits the solution
\begin{equation} \label{eq:fofvarphi}
 f(\varphi)=i\pi+\log\left(\frac{\sqrt{1-e^{4\varphi}}-1}{\sqrt{1-e^{4\varphi}}+1}\right)\equiv2\psi.
\end{equation}
It is then straightforward to show that $\psi$ satisfies
\begin{equation}
 \partial\bar{\partial}\psi=\frac{m^{2}}{2}\sinh(2\psi).
\end{equation}
To obtain (\ref{eq:result2}) we set
\begin{equation}
 e^{\Sigma_{\alpha}}=2kF(x,y)=e^{\chi}(\cosh(\varphi)+\sinh(\varphi))
\end{equation}
so that 
\begin{equation}
 \partial\bar{\partial}\Sigma_{\alpha}=\partial\bar{\partial}\chi+\partial\bar{\partial}\varphi.
\end{equation}
Using (\ref{eq:newward5.2}) and (\ref{eq:phi1}) it is possible to show that
\begin{equation}
  \partial\bar{\partial}\Sigma_{\alpha}=\frac{m^{2}}{2}(1-\cosh(2\psi))
\end{equation}
and so we have derived the known formula for the correlation functions (up to a constant factor which can be absorbed into the definitions of either the twist fields or the co-ordinates $z$ and $\bar{z}$).

Following our method we notice that it is now possible to parametrise $G(x,y) = \bra\sigma_{\alpha}(x,y)\sigma_{\alpha+1}(0)\ket$ in terms of a partial differential equation involving $\psi$. Setting
\begin{equation}
 e^{\Sigma'_{\alpha}}=2G(x,y)=e^{\chi}(\cosh(\varphi)-\sinh(\varphi))
\end{equation}
gives
\begin{align}
 \partial\bar{\partial}\Sigma'_{\alpha}= & -2\partial\varphi\bar{\partial}\varphi(1+\coth(2\varphi)) \nonumber \\
 = & \frac{2\tanh^{2}(\psi)}{\cosh(2\psi)-1}\partial\psi\bar{\partial}\psi.
\end{align}
This is a new result which is easily found following our construction. We also note that $H(x,y)$ can be written in terms of $\Sigma_{\alpha}$ and $\Sigma'_{\alpha}$, via the algebraic relation (\ref{eq:ward1}), and so it is also parametrised by the partial differential equations given above.


\section{Summary and perspectives} \label{sec:sum}

As stated in the introduction, we have derived Equations (\ref{eq:result1}), (\ref{eq:result2}), (\ref{eq:result3}) and (\ref{eq:result4}). One natural extension of this work is the derivation of a differential equation for 
\[
\langle\sigma_{\alpha}(x,y)\sigma_{\beta}(0,0)\rangle \quad \text{for}\quad \alpha\neq\beta, \beta\pm 1 
\]
as is presented (for $0<\alpha,\beta<1$) in \cite{BL97}. This calculation, however, is more complicated than that of the case $\alpha=\beta$, as some of the simplifications of Section \ref{sec:corfun} can not be applied. Some work is in progress in this direction \cite{Sprog}.

Further, the derivation presented here is very direct as it involves only the symmetries of the model together with the action of the conserved charge $Z$ on products of twist fields. It is also very general as it essentially only requires the density matrix to be an exponential of a quadratic form. It seems very likely that it could be applied to many other free field models possessing non-trivial twist fields. In particular, it would be very interesting to study, with such methods, twist fields associated to nonabelian symmetries.

Also, the techniques used to calculated form factors for the descendent twist fields could be applied in other free field models, or in other quantum states. In particular, this may lead to ways of calculating vacuum expectation values of primary twist fields more generally, in direct analogy to our calculation of $c_{\alpha}$. Besides being useful again in the study of twist fields associated to nonabelian symmetries, it could be applied for instance in conjunction with the notion of finite-temperature form factors developed in \cite{D05}, leading to expectation values at finite temperatures. 

Finally, a natural development is to determine the initial conditions that fix the solution to the differential equations corresponding to the correlation functions of interest. In particular, in analogy to what was done in \cite{DG08} following \cite{FZ03} for the Ising model, one should connect the finite-temperature form factors of $U(1)$ twist fields in the Dirac theory to the initial scattering data (or Jost function) associated to the sinh-Gordon equation.

\paragraph*{Acknowledgements}
J.S. would like to thank EPSRC for his studentship and E. Corrigan for asking the questions he should know the answer to.


\appendix

\section{Massless limit: CFT description and OPEs} \label{sec:cft}

\subsection{Bosonisation and leading OPEs}

If the mass is set to zero in the equations of motion (\ref{eom}), then the fields $\Psi_R$ and $\Psi_R^\dag$ become holomorphic (right-moving), and the fields $\Psi_L$ and $\Psi_L^\dag$ anti-holomorphic (left-moving). In this case, the non-zero two-point functions of Dirac fields become exactly
\beqa
	\bra\vac|{\cal T}\lt[\Psi_R^\dag(x_1,y_1)\Psi_R(x_2,y_2)\rt]|\vac\ket_{m=0} &=& \frc1{z_1-z_2},\n
	\bra\vac|{\cal T}\lt[\Psi_L^\dag(x_1,y_1)\Psi_L(x_2,y_2)\rt]|\vac\ket_{m=0} &=& \frc1{\b{z}_1-\b{z}_2}.
\eeqa
Multi-point functions are simply evaluated using Wick theorem. In particular, the right- and left-moving parts factorise.

As is well known, it is fruitful to reproduce these correlation functions using bosonisation. We introduce holomorphic and anti-holomorphic operators $e_\alpha(x,y)$ and $\b{e}_\alpha(x,y)$, $\alpha\in\R$. They have bosonic statistics and the following correlation functions:
\beqa
	\bra\vac|e_{\alpha_1}(x_1,y_1)\cdots e_{\alpha_n}(x_n,y_n)|\vac\ket_{m=0} &=& \delta_{0,\sum_j\alpha_j}
	\prod_{j<k} (-i(z_j-z_k))^{\alpha_j\alpha_k},\n
	\bra\vac|\b{e}_{\alpha_1}(x_1,y_1)\cdots \b{e}_{\alpha_n}(x_n,y_n)|\vac\ket_{m=0} &=& \delta_{0,\sum_j\alpha_j}
		\prod_{j<k} (i(\b{z}_j-\b{z}_k))^{\alpha_j\alpha_k}\label{corre}
\eeqa
for $y_1>y_2>\cdots>y_n$, where the power functions are on their principal branches. These correlation functions imply the following operator product expansions:
\beqa
	{\cal T}\lt[e_\alpha(x,y) e_{\alpha'}(x',y')\rt] &\sim & (-i(z-z'))^{\alpha\alpha'} \lt(1+\frc{\alpha}{\alpha+\alpha'} (z-z') \p'\rt) e_{\alpha+\alpha'}(x',y') \n
	{\cal T}\lt[\b{e}_\alpha(x,y) \b{e}_{\alpha'}(x',y')\rt] & \sim & (i(\b{z}-\b{z}'))^{\alpha\alpha'} \lt(1+\frc{\alpha}{\alpha+\alpha'} (\b{z}-\b{z}') \b\p'\rt) e_{\alpha+\alpha'}(x',y'),
	\label{OPEe}
\eeqa
as well as the following equal-time exchange relations for the operators $e_\alpha(x)$ and $\b{e}_\alpha(x)$:
\beqa \label{exe}
	e_{\alpha_1}(x_1) e_{\alpha_2}(x_2) &=&
		e^{i\pi\alpha_1\alpha_2\text{sgn}(x_2-x_1)} e_{\alpha_2}(x_2) e_{\alpha_1}(x_1), \n
	\b{e}_{\alpha_1}(x_1) \b{e}_{\alpha_2}(x_2) &=&
		e^{i\pi\alpha_1\alpha_2\text{sgn}(x_1-x_2)} e_{\alpha_2}(x_2) e_{\alpha_1}(x_1)
\eeqa
in both cases for $x_1\neq x_2$.

The basic principle of bosonisation is the equivalence of such correlation functions with those of the Dirac fields for appropriate choices of $\alpha$s. More precisely, with the identifications $\Psi_R = e^{-i\pi/4} \omega_R^{-1} e_{-1}$, $\Psi_R^\dag = e^{-i\pi/4} \omega_R e_1$, $\Psi_L = e^{i\pi/4} \omega_L e_{1}$ and  $\Psi_L^\dag = e^{i\pi/4} \omega^{-1}_L e_{-1}$, where $\omega_R$ and $\omega_L$ are pure phases, the correlation functions involving $\Psi_R$, $\Psi_R^\dag$ are reproduced, and those involving $\Psi_L$, $\Psi_L^\dag$ as well. In particular, the exchange relations (\ref{exe}) reproduce the fermionic statistics for exchanges in the right- and left-moving sector separately. However, in order to reproduce the correlation functions involving a mixture of right- and left-moving Dirac fields, we need to introduce extra elements guaranteeing fermionic exchange relations amongst right- and left-movers: so-called Klein factors. For this purpose, we will use quaternionic elements $\qi$, $\qj$ and $\qk$, with both $\Psi_R$ and $\Psi_R^\dag$ proportional to $\qj$, and both $\Psi_L$ and $\Psi_L^\dag$ proportional to $\qi$. In any given sectors, such quaternionic elements can be seen as being part of the pure phases $\omega_{R,L}$, hence do not affect the correlation functions (using $\qi^{-1} = -\qi$, etc.). For correlation functions of mixed sectors, the anti-commutation $\qi \qj = -\qj \qi$ guarantees the correct statistics of Dirac fields. Since only even numbers of right- and left-moving Dirac fields are involved in non-zero correlation functions, the quaternionic element $\qk$ never remains. The precise identification can then be written as follows:
\beq\label{relpsie}
	\Psi_R = -e^{-i\pi/4}\omega^{-1}\qj e_{-1},\quad \Psi_R^\dag = e^{-i\pi/4}\omega\qj e_1,\quad
	\Psi_L = e^{i\pi/4}\omega\qi \b{e}_{1},\quad \Psi_L^\dag = -e^{i\pi/4}\omega^{-1}\qi \b{e}_{-1}.
\eeq
Here, we have used the $U(1)$ invariance of the Dirac theory in order to remain with only one arbitrary phase $\omega$.

The bosonisation relations allow us to put in the same framework the $U(1)$-twist fields discussed in Section \ref{sec:model}. The basis for the identification is the comparison between the exchange relations (\ref{eq:equaltimebraiding}) and (\ref{exe}). Denoting by $Q$ the $U(1)$-charge of the Dirac theory, with $[Q,\Psi_{R,L}^\dag] = -\Psi_{R,L}^\dag$, $[Q,\Psi_{R,L}] = \Psi_{R,L}$, we can simply make the identification:
\beq\label{sige}
	\sigma_\alpha = e_\alpha \b{e}_\alpha e^{-i\pi\alpha Q}.
\eeq
The action of $Q$ on the operators $e_\alpha$ and $\b{e}_\alpha$ is given by
\beq\label{Qe}
	e^{icQ}e_\alpha e^{-icQ} = e^{-ic\alpha}e_\alpha,\quad e^{icQ}\b{e}_\alpha e^{-icQ} = e^{ic\alpha}\b{e}_\alpha.
\eeq
Note that relation (\ref{sige}) along with the correlation functions (\ref{corre}) are in agreement with the normalisation (\ref{normsigm}).

Correlation functions (\ref{corre}) can be realised using the Heisenberg algebra: one usually write $e_\alpha(x,y) = e^{i\alpha\varphi_R(x,y)}$ and $\b{e}_\alpha(x,y) = e^{-i\alpha\varphi_L(x,y)}$ where $\varphi_{R,L}$ are (appropriately normalised) holomorphic and anti-holomorphic free massless bosonic field, respectively. The $U(1)$ symmetry in terms of such fields corresponds to a shift of both $\varphi_R$ and $\varphi_L$ by the same constant, and the arbitrariness of the phase $\omega$ in (\ref{relpsie}) can be thought of as an arbitrariness under shifts of $\varphi_{R}$ and $\varphi_L$ by opposite constants. Also, in terms of the bosonic fields, the $U(1)$-twist fields is simply $\sigma_\alpha = e^{i\alpha\varphi}e^{-i\pi\alpha Q}$, with $\varphi := \varphi_R - \varphi_L$.

From relation (\ref{sige}) along with the correlation functions (\ref{corre}), it is possible to determine the leading coefficients of operator product expansions (OPEs) involving Dirac fields and the $U(1)$ twist field. Taking into consideration the exchange relations (\ref{eq:equaltimebraiding}) and the correlation functions (\ref{corre}), we find
\beqa
 {\cal T} \lt[ \Psi_{R}^{\dagger}(x,y)\sigma_{\alpha}(0) \rt] &\sim&
 	(-iz)^{\alpha} \sigma_{\alpha+1,\alpha}(0), \n
 {\cal T}\lt[\Psi_{R}(x,y)\sigma_{\alpha}(0)\rt]&\sim&(-iz)^{-\alpha}\sigma_{\alpha-1,\alpha}(0), \label{OPECFTR}
\eeqa
where we define, for $\alpha\in\R$,
\beqa
	\sigma_{\alpha+1,\alpha} &:=& e^{-i\pi/4}\omega\qj e_{\alpha+1}\b{e}_\alpha e^{-i\pi \alpha Q}, \n
	\sigma_{\alpha-1,\alpha} &:=& -e^{-i\pi/4}\omega^{-1}\qj e_{\alpha-1}\b{e}_\alpha e^{-i\pi \alpha Q}. \label{defspm1}
\eeqa
The power functions on the right-hand sides in (\ref{OPECFTR}) are on their principal branches; note that there is agreement between the phases occurring from exchange relations (\ref{eq:equaltimebraiding}) and the phase differences occurring through the cuts of the principal branches. Further, we also find
\beqa
 {\cal T}\lt[\Psi_{L}^{\dagger}(x,y)\sigma_{\alpha}(0)\rt]&\sim&
 	-i\omega^{-2} \qk e^{-i\pi Q} (i\bar{z})^{-\alpha} \sigma_{\alpha,\alpha-1}(0) \n
 {\cal T}\lt[\Psi_{L}(x,y)\sigma_{\alpha}(0)\rt]&\sim& -i \omega^2 \qk e^{i\pi Q} (i\bar{z})^{\alpha}\sigma_{\alpha,\alpha+1}(0).
 \label{OPECFTLinc}
\eeqa
Note that relation (\ref{ts}) can readily be obtained from (\ref{relpsie}), (\ref{defspm1}) and (\ref{sige}). However, in order to simplify the OPEs (\ref{OPECFTLinc}), we need to specify both $\qk$ and $\omega$.

The CFT has both the $U(1)$ symmetry, and a symmetry under constant shifts of $\varphi$. Hence, there does not seem to be any fundamental way of determining the phase $\omega$ in the CFT context. There are non-zero correlation functions involving odd numbers of both right- and left-movers, if twist fields are also inserted. Such correlation functions will involve remaining $\qk$ factors, as well as remaining phases $\omega$. Although fixing these correlation functions would partly lift these ambiguities, there is no clear principle as to how to fix them.

In the massive theory, there is an (almost) unambiguous way of writing the element $\qk$ explicitly as an operator. Indeed, $\qk$ is an operator that anti-commutes with all Dirac fields, that commutes with the twist field $\sigma_\alpha$, and that squares to $-1$, so we may simply write it as
\beq\label{qk}
	\qk = \ep \,i\,e^{i\pi Q}
\eeq
for some sign $\ep=\pm$. In particular, we have $\bra\vac|\qk|\vac\ket = \ep\,i$. It is important that we can fix $\qk$, because there are unambiguous non-zero correlation functions involving odd numbers of both (massively deformed) right- and left-movers; in the bosonised language, these involve a remaining operator $\qk$, which needs to be well-defined. We note that in the present calculations, we do not need to determine the sign $\ep$.

In addition to being able to fix $\qk$, in the massive theory, the symmetry under constant shifts of $\varphi$ is broken. This symmetry breaking should allow us to evaluate the phase $\omega$. The symmetry breaking is simple to see by writing explicitly the mass term in the Hamiltonian:
\[
	\frc{im}{4\pi} \int dx\,\lt(\Psi_R^\dag(x) \Psi_L(x) -\Psi_L^\dag(x)\Psi_R(x)\rt) =
		-\frc{im\qk}{2\pi}\int dx\, \cos(\varphi(x)+2\nu)
\]
with $\omega =: e^{i\nu}$. In order to determine $\omega$, we simply define $\varphi$ by specifying that the twist field $\sigma_\alpha$ has real and positive vacuum expectation value. Then, we use the fact that the field $e^{i\varphi}$ has real and positive vacuum expectation value if and only if the integrand in the mass term has minima at $\varphi=0\ {\rm mod}\ 2\pi$ (so that the main contributions, in a path integral perspective, are at values of $\varphi$ symmetrically distributed around $0$: values of $e^{i\varphi}$ add up to a positive real number). Hence, the integrand in the mass term must be $\cos\varphi$ and the coefficient in front of the integral must be negative in bosonic sectors (i.e. replacing $\qk$ by $\bra\vac|\qk|\vac\ket$). Therefore, the requirement is that
\beq\label{repom}
	\omega^2 = -\ep.
\eeq
Combining (\ref{qk}), (\ref{repom}) with (\ref{OPECFTLinc}), we find, independently of $\ep$, the OPEs
\beqa
 {\cal T}\lt[\Psi_{L}^{\dagger}(x,y)\sigma_{\alpha}(0)\rt]&\sim&
 	-(i\bar{z})^{-\alpha} \sigma_{\alpha,\alpha-1}(0) \n
 {\cal T}\lt[\Psi_{L}(x,y)\sigma_{\alpha}(0)\rt]&\sim& -(i\bar{z})^{\alpha}\sigma_{\alpha,\alpha+1}(0).
 \label{OPECFTL}
\eeqa

It is a simple matter to use the definitions and relations found above in order to define other non-derivative descendent fields of the Dirac theory, and to establish the leading and first sub-leading terms in the OPEs with the Fermion fields. We define recursively $\sigma_{\alpha\pm n,\alpha}$ for integer $n\ge 0$ via
\beqa
 {\cal T} \lt[ \Psi_{R}^{\dagger}(x,y)\sigma_{\alpha+n,\alpha}(0) \rt] &\sim&
 	(-iz)^{\alpha+n} \sigma_{\alpha+n+1,\alpha}(0), \n
 {\cal T}\lt[\Psi_{R}(x,y)\sigma_{\alpha-n,\alpha}(0)\rt]&\sim&(-iz)^{n-\alpha}\sigma_{\alpha-n-1,\alpha}(0).
\eeqa
This gives the expressions, for $\alpha\in\R$ and $n\ge0$ integer,
\beqa
	\sigma_{\alpha+n,\alpha} &:=& (e^{-i\pi/4}\omega\qj)^n e_{\alpha+n}\b{e}_\alpha e^{-i\pi \alpha Q}, \n
	\sigma_{\alpha-n,\alpha} &:=& (-e^{-i\pi/4}\omega^{-1}\qj)^n e_{\alpha-n}\b{e}_\alpha e^{-i\pi \alpha Q}
\eeqa
(these are in agreement with (\ref{sige}) and (\ref{defspm1}) in the cases $n=0$ and $n=1$ respectively). From these expressions along with the rules (\ref{relpsie}), (\ref{qk}), (\ref{repom}) and (\ref{Qe}), and with the OPEs (\ref{OPEe}), we find the OPEs (for $\alpha-\beta\in\Z$):
\beqa
	{\cal T}\lt[\Psi_R^\dag(x,y)\sigma_{\alpha,\beta}(0) \rt] &\sim & (-i)^{\Theta(\beta-\alpha-1/2)} (-iz)^\alpha \lt(1+\frc{1}{1+\alpha} z\p \rt) \sigma_{\alpha+1,\beta}(0) \n
	{\cal T}\lt[\Psi_R(x,y)\sigma_{\alpha,\beta}(0) \rt] &\sim & (-i)^{\Theta(\alpha-\beta-1/2)} (-iz)^{-\alpha} \lt(1+\frc{1}{1-\alpha} z\p \rt) \sigma_{\alpha-1,\beta}(0) \n
	{\cal T}\lt[\Psi_L^\dag(x,y)\sigma_{\alpha,\beta}(0) \rt] &\sim & (-i)^{\Theta(\beta-\alpha-1/2)} (-1)^{\alpha-\beta+1} (i\b{z})^{-\beta} \lt(1+\frc{1}{1-\beta} \b{z}\b\p \rt) \sigma_{\alpha,\beta-1}(0) \n
	{\cal T}\lt[\Psi_L(x,y)\sigma_{\alpha,\beta}(0) \rt] &\sim & (-i)^{\Theta(\alpha-\beta-1/2)} (-1)^{\alpha-\beta+1}(i\b{z})^{\beta} \lt(1+\frc{1}{1+\beta} \b{z}\b\p \rt) \sigma_{\alpha,\beta+1}(0) \label{OPEPsiSig}
\eeqa
where $\Theta$ is the Heaviside step-function, $\Theta(\gamma)=1\ (\gamma>0),\ 0\ (\gamma<0)$, and the derivatives on the right-hand side are with respect to the arguments of the twist fields, which are afterwards put to 0. In these OPEs, the next correction term is equivalent to a term of order $z^2$ (for the first two) or $\b{z}^2$ (for the last two) inside the large parenthesis (but generally involve non-derivative descendents of the twist fields written).

We may write these OPEs more generally by putting the twist fields at position $x',y'$ and by replacing $z$ by $z-z'$ and $\b{z}$ by $\b{z}-\b{z}'$ in the power functions on the right-hand side. From this, taking derivatives with respect to $z'$ and $\b{z}'$, we obtain OPE's with first-derivatives of twist fields:
\beqa
	{\cal T}\lt[\Psi_R^\dag(x,y)\p\sigma_{\alpha,\beta}(0) \rt] &=& (-i)^{\Theta(\beta-\alpha-1/2)} (i\alpha) (-iz)^{\alpha-1}  \sigma_{\alpha+1,\beta}(0) + O(z^{\alpha+2}) \n
	{\cal T}\lt[\Psi_R(x,y)\p\sigma_{\alpha,\beta}(0) \rt] &= & (-i)^{\Theta(\alpha-\beta-1/2)} (-i\alpha) (-iz)^{-\alpha-1}  \sigma_{\alpha-1,\beta}(0) + O(z^{-\alpha+2})\n
	{\cal T}\lt[\Psi_L^\dag(x,y)\b\p\sigma_{\alpha,\beta}(0) \rt] &= & (-i)^{\Theta(\beta-\alpha-1/2)} (-1)^{\alpha-\beta+1} (i\beta) (i\b{z})^{-\beta-1}  \sigma_{\alpha,\beta-1}(0) + O(\b{z}^{-\beta+2})\n
	{\cal T}\lt[\Psi_L(x,y)\b\p\sigma_{\alpha,\beta}(0) \rt] &= & (-i)^{\Theta(\alpha-\beta-1/2)} (-1)^{\alpha-\beta+1}(-i\beta) (i\b{z})^{\beta-1} \sigma_{\alpha,\beta+1}(0)+ O(\b{z}^{\beta+2}). \n \label{OPEPsiSigDer}
\eeqa
Note that the first non-trivial corrections to the leading OPEs, corresponding to pure derivative of twist fields, are exactly zero.


\subsection{Action of the charge $Z$}

The OPEs above along with the integral expression (\ref{eq:defZ}) of the charge $Z$ allow us to evaluate the action of $Z$ on the product of twist fields belonging to different copies as follows. First, in the CFT the charge $Z$ decomposes into a sum of two conserved charges, $Z=Z_R+Z_L$, where $Z_{R,L}$ involves only right-moving and left-moving Fermions respectively. Second, by the usual principles of QFT, commutators of local charges with local fields can be evaluated through contour integrals and the residue theorem. Using the replacements $-iz\mapsto s$ and $i\b{z}\mapsto s$ respectively, the commutators with $Z_{R}$ and $Z_{L}$ are recast into the following contour integrals:
\beqa
	[Z_R,\Or(0)] &\mapsto& \frc1{2\pi} \oint ds \lt(\Psi_R(x,y)\Phi_R^\dag(x,y) + \Psi_R^\dag(x,y)\Phi_R(x,y)\rt) \Or(0) \n
{}	[Z_L,\Or(0)] &\mapsto& -\frc1{2\pi} \oint ds \lt(\Psi_L(x,y)\Phi_L^\dag(x,y) + \Psi_L^\dag(x,y)\Phi_L(x,y)\rt) \Or(0) \no.
\eeqa
The OPEs (\ref{OPEPsiSig}) above immediately give rise to Equations (\ref{eq:1stzrels}), and the OPEs (\ref{OPEPsiSigDer}) provide actions of $Z$ on products of twist fields involving one derivative that agree with those written in Section \ref{sec:eqZ}.


\section{Functional equations without translation and parity symmetry} \label{sec:eq-rtp}

In this appendix we write down the differential equations arising from the Ward Identities (\ref{eq:WardIdents}) in the absence of translation and parity symmetry. The loss of these symmetries means that we no longer have all the nice symmetry properties discussed in section \ref{sec:syms}. As a result we must introduce a more appropriate, but cumbersome, notation. We define
\beq
F_{\alpha,\beta}^{\gamma,\delta}(x_{1},y_{1},x_{2},y_{2}):=\bra\sigma_{\alpha,\beta}(x_{1},y_{1})\sigma_{\gamma,\delta}(x_{2},y_{2})\ket
\eeq
and first note that the lack of translation symmetry prevents us from always placing the second field at $(0,0)$, and so the correlation functions must remain functions of four variables. It should also be noted that
\begin{eqnarray}
 F_{\alpha,\alpha}^{\beta,\beta}(x_{1},y_{1},x_{2},y_{2})=\langle\sigma_{\alpha,\alpha}(x_{1},y_{1})\sigma_{\beta,\beta}(x_{2},y_{2})\rangle & =\langle\sigma_{\beta,\beta}(x_{2},y_{2})\sigma_{\alpha,\alpha}(x_{1},y_{1}) \nonumber \\ & \rangle\neq F_{\beta,\beta}^{\alpha,\alpha}(x_{1},y_{1},x_{2},y_{2})
\end{eqnarray}
and so the Ward identities can not be expected to simplify as much as they did in the previous case.

The Ward identities in this case are
\begin{subequations} \label{eq:WardIdentsB}
 \begin{align}
  \langle[Z,\sigma_{\alpha,\alpha}^{\Psi}(x_{1},y_{1})\sigma_{\alpha+1,\alpha+1}^{\Phi}(x_{1},y_{1})\sigma_{\alpha+1,\alpha}^{\Psi}(x_{2},y_{2})\sigma_{\alpha,\alpha+1}^{\Phi}(x_{2},y_{2})]\rangle=0  \\
  \langle[Z,[P,\sigma_{\alpha,\alpha}^{\Psi}(x_{1},y_{1})\sigma_{\alpha+1,\alpha+1}^{\Phi}(x_{1},y_{1})]\sigma_{\alpha+1,\alpha}^{\Psi}(x_{2},y_{2})\sigma_{\alpha,\alpha+1}^{\Phi}(x_{2},y_{2})]\rangle=0  \\
  \langle[Z,[\bar{P},\sigma_{\alpha,\alpha}^{\Psi}(x_{1},y_{1})\sigma_{\alpha+1,\alpha+1}^{\Phi}(x_{1},y_{1})]\sigma_{\alpha+1,\alpha}^{\Psi}(x_{2},y_{2})\sigma_{\alpha,\alpha+1}^{\Phi}(x_{2},y_{2})]\rangle=0 \\
  \langle[Z,[P,[\bar{P},\sigma_{\alpha,\alpha}^{\Psi}(x_{1},y_{1})\sigma_{\alpha+1,\alpha+1}^{\Phi}(x_{1},y_{1})]]\sigma_{\alpha,\alpha+1}^{\Psi}(x_{2},y_{2})\sigma_{\alpha+1,\alpha}^{\Phi}(x_{2},y_{2})]\rangle=0  \\
  \langle[Z,[P,\sigma_{\alpha,\alpha}^{\Psi}(x_{1},y_{1})\sigma_{\alpha+1,\alpha+1}^{\Phi}(x_{1},y_{1})][\bar{P},\sigma_{\alpha,\alpha+1}^{\Psi}(x_{2},y_{2})\sigma_{\alpha+1,\alpha}^{\Phi}(x_{2},y_{2})]]\rangle=0
 \end{align}
\end{subequations}
and their expansions in terms of correlation functions are cumbersome. To ease our notation the dependence on $(x_{1},y_{1},x_{2},y_{2})$ will be implicit in all $F_{\alpha,\beta}^{\gamma,\delta}$, as operators at $(x_{1},y_{1})$ are always to the left of those at $(x_{2},y_{2})$. We also introduce the symmetric, bilinear differential operator
\begin{equation}
 \mathcal{D}_{i,j}(u,v)=u\partial_{i}\bar{\partial}_{j}v+v\partial_{i}\bar{\partial}_{j}u-\partial_{i} u\bar{\partial}_{j}v-\bar{\partial}_{j}u\partial_{i} v
\end{equation}
where $z_{j}=\frac{-i}{2}(x_{j}+iy_{j})$ and
\begin{eqnarray}
 \partial_{i}=\frac{\partial}{\partial z_{i}} & \text{and} & \bar{\partial}_{i}=\frac{\partial}{\partial \bar{z}_{i}}.       
\end{eqnarray}

Now we can write (\ref{eq:WardIdentsB}) as 
\begin{subequations}
 \begin{equation}
  F_{\alpha,\alpha+1}^{\alpha+1,\alpha}F_{\alpha+1,\alpha}^{\alpha,\alpha+1}-F_{\alpha,\alpha}^{\alpha+1,\alpha+1}F_{\alpha+1,\alpha+1}^{\alpha,\alpha}+F_{\alpha,\alpha}^{\alpha,\alpha}F_{\alpha+1,\alpha+1}^{\alpha+1,\alpha+1}=0
 \end{equation}
 \begin{multline}
  \partial_{z_{1}}F_{\alpha,\alpha+1}^{\alpha+1,\alpha}F_{\alpha+1,\alpha}^{\alpha,\alpha+1}-F_{\alpha,\alpha+1}^{\alpha+1,\alpha}\partial_{z_{1}}F_{\alpha+1,\alpha}^{\alpha,\alpha+1} +\partial_{z_{1}}F_{\alpha,\alpha}^{\alpha+1,\alpha+1}F_{\alpha+1,\alpha+1}^{\alpha,\alpha}-F_{\alpha,\alpha}^{\alpha+1,\alpha+1}\partial_{z_{1}}F_{\alpha+1,\alpha+1}^{\alpha,\alpha} \\ -\partial_{z_{1}}F_{\alpha,\alpha}^{\alpha,\alpha}F_{\alpha+1,\alpha+1}^{\alpha+1,\alpha+1}+F_{\alpha,\alpha}^{\alpha,\alpha}\partial_{z_{1}}F_{\alpha+1,\alpha+1}^{\alpha+1,\alpha+1}=0
 \end{multline}
 \begin{multline}
  \bar{\partial}_{z_{1}}F_{\alpha,\alpha+1}^{\alpha+1,\alpha}F_{\alpha+1,\alpha}^{\alpha,\alpha+1}-F_{\alpha,\alpha+1}^{\alpha+1,\alpha}\bar{\partial}_{z_{1}}F_{\alpha+1,\alpha}^{\alpha,\alpha+1} +\bar{\partial}_{z_{1}}F_{\alpha+1,\alpha+1}^{\alpha,\alpha}F_{\alpha,\alpha}^{\alpha+1,\alpha+1}-F_{\alpha+1,\alpha+1}^{\alpha,\alpha}\bar{\partial}_{z_{1}}F_{\alpha,\alpha}^{\alpha+1,\alpha+1} \\ -\bar{\partial}_{z_{1}}F_{\alpha+1,\alpha+1}^{\alpha+1,\alpha+1}F_{\alpha,\alpha}^{\alpha,\alpha}+F_{\alpha+1,\alpha+1}^{\alpha+1,\alpha+1}\bar{\partial}_{z_{1}}F_{\alpha,\alpha}^{\alpha,\alpha}=0
 \end{multline}
 \begin{multline}
  \mathcal{D}_{1,1}(F_{\alpha+1,\alpha}^{\alpha,\alpha+1},F_{\alpha,\alpha+1}^{\alpha+1,\alpha})-m^{2}F_{\alpha+1,\alpha}^{\alpha,\alpha+1}F_{\alpha,\alpha+1}^{\alpha+1,\alpha} \\ -\mathcal{D}_{1,1}(F_{\alpha,\alpha}^{\alpha,\alpha},F_{\alpha+1,\alpha+1}^{\alpha+1,\alpha+1})+\mathcal{D}_{1,1}(F_{\alpha,\alpha}^{\alpha+1,\alpha+1},F_{\alpha+1,\alpha+1}^{\alpha,\alpha})=0
 \end{multline}
 \begin{equation}
  \mathcal{D}_{1,2}(F_{\alpha,\alpha+1}^{\alpha+1,\alpha},F_{\alpha+1,\alpha}^{\alpha,\alpha+1})-\mathcal{D}_{1,2}(F_{\alpha,\alpha}^{\alpha,\alpha},F_{\alpha+1,\alpha+1}^{\alpha+1,\alpha+1})-\mathcal{D}_{1,2}(F_{\alpha,\alpha}^{\alpha+1,\alpha+1},F_{\alpha+1,\alpha+1}^{\alpha,\alpha})=0.
 \end{equation}
\end{subequations}
It should be noted that this system is under defined as there are $6$ independent functions, each of two variables, and only $5$ equations. It is possible to find more Ward identities leading to non-trivial relations between the correlation functions but as we are unable to identify any integrable structure in these equations we only include the above equations for completeness.

\end{document}